\documentclass[pre,showpacs,10pt,a4paper]{revtex4}

\usepackage{graphicx}
\usepackage{epsfig}
\usepackage{amssymb}
\usepackage{amsfonts}
\usepackage{amsmath}

\begin{document}

\title{Renormalized cluster expansion of the microfield distribution in a
strongly coupled two-component plasmas}
\author{H.B.Nersisyan}
\email{hrachya@irphe.am}
\author{D.A.Osipyan}
\affiliation{Division of Theoretical Physics, Institute of Radiophysics and Electronics,
Alikhanian Brothers str.1, 378410 Ashtarak, Armenia}
\author{G.Zwicknagel}
\email{zwicknagel@theorie2.physik.uni-erlangen.de}
\affiliation{Institut f\"{u}r Theoretische Physik II, Erlangen-N\"{u}rnberg Universit\"{a}%
t, Staudtstrasse 7, D-91058 Erlangen, Germany}
\date{\today}

\begin{abstract}
The electric microfield distribution (MFD) at an impurity ion is studied for
two-component (TCP) electron-ion plasmas using molecular dynamics simulation
and theoretical models. The particles are treated within classical
statistical mechanics using an electron-ion Coulomb potential regularized at
distances less than the de Broglie length to take into account
quantum-diffraction effects. Corrections to the potential-of-mean-force
exponential (PMFEX) approximation recently proposed for MFD in a strongly
coupled TCP [Phys. Rev. E \textbf{72}, 036403 (2005)] are obtained and
discussed. This has been done by a generalization of the standard
Baranger-Mozer and renormalized cluster expansion techniques originally
developed for the one-component plasmas to the TCPs. The results obtained
for a neutral point are compared with those from molecular dynamics
simulations. It is shown that the corrections do not help to improve the
PMFEX approximation for a TCP with low ionic charge $Z$. But starting
with $Z > 5$ the PMFEX model is substantially improved and the agreement
with numerical simulations is excellent. We have also found that with increasing
coupling the PMFEX approximation becomes invalid to predict the MFD at a neutral 
point while its corrected version agrees satisfactory with the simulations.
\end{abstract}

\maketitle

\section{Introduction}
\label{sec:int}

The determination of the electric microfield distribution at a neutral atom
or charged impurity ion (radiator) in a plasma is an important tool for the
understanding of many spectroscopic experiments \cite{gri97,sal98}. Since
the pioneering work of Holtsmark \cite{hol19}, who completely neglected
correlations between the particles (ideal plasma), many efforts have been
concentrated on an improved statistical description of the microfield
distribution. The first theory which goes beyond the Holtsmark limit and
which is based on a cluster expansion similar to that of Ursell and Mayer 
\cite{may40} was developed by Baranger and Mozer \cite{bar59,moz60}. In this
approach the microfield distribution is represented as an expansion in terms
of correlation functions which has been truncated on the level of the pair
correlation. The latter is treated in the Debye-H\"{u}ckel form which
corresponds to the first order of the expansion in the coupling parameter.
The theory by Baranger and Mozer was improved by Hooper \cite{hoo66,hoo68}.
They reformulated the expansion of the microfield distribution in terms of
other functions by introducing a free parameter which was adjusted in such a
way to arrive at a level where the resulting microfield distribution did not
depend on the free parameter any more. However, it was argued that such a
method is only valid at small coupling parameters, where the correction to
the Holtsmark distribution, corresponding to the first term in the series,
is small. The first theory capable to provide reliable numerical results for
strongly coupled plasmas, known as adjustable-parameter exponential
approximation (APEX), was proposed by Iglesias and co-workers \cite%
{igl83,ala84}. It involves a noninteracting quasiparticle representation of
the electron-screened ions. This phenomenological but highly successful
approximation is based on a parameterization of the electric microfield
produced on a radiator with the Debye-H\"{u}ckel-type screened interaction
with unknown screening length. This free parameter is then adjusted in such
a way to yield the correct second moment of the microfield distribution.
Afterwards the APEX model was substantially improved for neutral radiators
using renormalized correlation functions and electric fields \cite{duf85}.
(We refer the reader to Ref.~\cite{rom06} for a recent review).

Most of this work was done on one-component plasmas (OCP) and thus neglects the
influence of the attractive interactions between electrons and ions. In this
paper we study the microfield distribution in
a two-component plasmas (TCP). This has been done previously in Refs.~\cite%
{yan86,ort00} for partially degenerate electrons. In Ref.~\cite{yan86} the
low-frequency component of the microfield was calculated within the linear
response treatment taking strong correlations into account via local field
corrections. Also the problem of attractive interaction has been considered
for a single but highly charged impurity ion immersed in an electronic OCP 
\cite{tal02}. In the recent papers \cite{ner05,ner06} we already studied
strongly coupled systems, i.e., a highly charged radiator in a TCP of
classical (nondegenerated) and strongly correlated particles beyond a
perturbative treatment. As in Ref.~\cite{yan86} the theoretical scheme
(PMFEX) presented in Refs.~\cite{ner05,ner06} is based on the
potential-of-mean-force approximation. It exactly satisfies the sum-rule
requirement arising from the second moment of the microfield distribution
without introducing adjustable parameters. Comparisons of the PMFEX
calculations with the molecular-dynamics simulations for the electric fields
at the highly charged points relevant for laser-produced plasmas show, in
general, an excellent agreement even for large coupling \cite{ner05}.
However, the results of the PMFEX for neutral points have not yet been
comprehensively tested. Here, the PMFEX scheme may be less accurate \cite{ner06}, similar
as it occurs in the APEX where further improvement for microfield
distribution at neutral point is needed, see Ref.~\cite{duf85}. Moreover, in a regime
dominated by small local fields and hence by small local electronic density
the PMFEX deviates from the molecular-dynamics simulations also for charged
radiator \cite{ner05}. This feature has been clearly observed for a single
ion embedded in an electronic OCP in Ref.~\cite{tal02}. Therefore, it is of
interest to improve the PMFEX model and provide a tool for calculating the
corrections to it when required. Our purpose here is to introduce the key
ideas of PMFEX in the standard Baranger-Mozer cluster representation for the
microfield distributions \cite{bar59,moz60}. In this way some contact
between PMFEX and standard small plasma-parameter theories is established,
while also providing PMFEX as the leading term in a series from which
corrections can be calculated explicitly. Another important ingredient of
our treatment is the electron-ion attractive interaction which drastically
changes the physical properties of the system as compared to classical OCPs
(see, e.g., Ref.~\cite{tal02}). But the thermodynamic stability of a TCP
requires at least some quantum features for the electron-ion interaction at
short distances. This required minimum of quantum features is here taken
into account by using a regularized ion-electron potential \cite%
{mor59,kel63,deu81}, which then enables the application of classical
statistical mechanics and classical MD simulations.

The paper is organized as follows. In Sec.~\ref{sec:1}, we define the basic
parameters of interest for a TCP. The theoretical schemes applied previously
to either electronic or ionic OCPs are generalized to TCPs in Secs.~\ref%
{sec:2} and \ref{sec:3}. In particular, starting with the canonical ensemble
formulation of the microfield distribution in the TCP the Baranger-Mozer
treatment is reviewed and the structure of the resulting cluster series is
discussed in Sec.~\ref{sec:2}. Next, the basic assumptions of the PMFEX are
briefly given and motivated in Sec.~\ref{sec:3}. Here a formal relationship
of the PMFEX to the Baranger-Mozer series is established and the first two
terms of a renormalized cluster series are given explicitly. Furthermore, to
test obtained theoretical results in Sec.~\ref{sec:4} the correction terms
are calculated at a neutral point and compared with the results from
classical molecular dynamics simulations. The results are summarized in Sec.~%
\ref{sec:sum}. The exact second moment of the microfield distribution at a
neutral point is considered in the Appendix \ref{sec:app1}.

\section{Basic parameters for the TCP}
\label{sec:1}

We consider a neutral and isotropic two-component electron-ion plasma
consisting of ions and electrons at a temperature $T$ in a volume $\Omega $.
The particles are assumed to be classical and pointlike. The average
densities, charges, and masses of the ions and electrons are $n_{i}$, $n_{e}$%
, and $Ze$, $-e$ and $m_{i}$, $m$, respectively. We assume that the density
of radiator ions or atoms are small, $n_{R}\ll n_{i;e}$ and thus consider
only one radiator ion with charge $Z_{R}e$ in our calculations (throughout
this paper the index $R$ refers to the radiators). Because of the charge
neutrality we have $n_{e}=n_{i}Z$.

We now introduce the Coulomb coupling parameters $\Gamma _{\alpha \beta }$.
Introducing the Wigner-Seitz radii, i.e., the mean electron-electron,
electron-ion, and ion-ion distances through the relations, $a_{e}^{-3}=4\pi
n_{e}/3$, $a^{-3}=4\pi n/3$, and $a_{i}^{-3}=4\pi n_{i}/3$ (where $%
n=n_{e}+n_{i}$ is the plasma total density) these parameters are defined as 
\begin{equation}
\Gamma _{ee}=\frac{e_{S}^{2}}{a_{e}k_{B}T},\quad \Gamma _{ei}=\frac{%
Ze_{S}^{2}}{ak_{B}T},\quad \Gamma _{ii}=\frac{Z^{2}e_{S}^{2}}{a_{i}k_{B}T},
\label{eq:1}
\end{equation}%
respectively, where $e_{S}^{2}=e^{2}/4\pi \varepsilon _{0}$. Note that 
\begin{equation}
\Gamma _{ee}=\frac{\Gamma _{ei}}{\left[ Z^{2}\left( Z+1\right) \right] ^{1/3}%
},\quad \Gamma _{ii}=\frac{Z\Gamma _{ei}}{\left( Z+1\right) ^{1/3}}.
\label{eq:2}
\end{equation}%
In a hydrogen plasma with $Z=1$ we obtain $\Gamma _{ee}=\Gamma
_{ii}=2^{-1/3}\Gamma _{ei}$ while in a plasma with highly charged ions ($%
Z\gg 1$) $\Gamma _{ii}=Z^{2/3}\Gamma _{ei}\gg \Gamma _{ee}=\Gamma _{ei}/Z$.
So, in the TCP with uncorrelated electrons the ions may be strongly
correlated.

The Holtsmark field $E_{H}$ for a TCP is given by $%
E_{H}^{3/2}=E_{He}^{3/2}+E_{Hi}^{3/2}$ (see Ref.~\cite{ner05} for details),
where $E_{He}$ and $E_{Hi}$ are the electronic and ionic Holtsmark fields,
respectively, $E_{He}=Ce_{F}/a_{e}^{2}$, $E_{Hi}=CZe_{F}/a_{i}^{2}$ with $%
C=\left( 8\pi /25\right) ^{1/3}$ and $e_{F}=e/4\pi \varepsilon _{0}$. Since $%
E_{He}=Z^{-1/3}E_{Hi}$ the electronic and ionic components of a hydrogen TCP
contribute equally to the Holtsmark field. For a completely ionized TCP with
highly charged ions the ions dominate $E_{H}$. The definition of the
Holtsmark field $E_{H}$ for a TCP is equivalent to the obvious relation $%
n=n_{e}+n_{i}$ and can be represented as 
\begin{equation}
E_{H}=\frac{C\mathcal{Z}e_{F}}{a^{2}},\qquad \mathcal{Z}=\left[ \frac{%
Z\left( 1+Z^{1/2}\right) }{\left( Z+1\right) }\right] ^{2/3}  \label{eq:H}
\end{equation}%
with an effective charge $\mathcal{Z}$.

For the regularized pair interaction potential $e_{S}^{2}q_{\alpha }q_{\beta
}u_{\alpha \beta }\left( r\right) $ with $\alpha ,\beta =e,i,R$, $q_{e}=-1$, 
$q_{i}=Z$, $q_{R}=Z_{R}$, we here employ 
\begin{equation}
u_{\alpha \beta }\left( r\right) =\frac{1}{r}\left( 1-e^{-r/\delta _{\alpha
\beta }}\right) .  \label{eq:3}
\end{equation}%
The cutoff parameters $\delta _{\alpha \beta }$ are related to the thermal
de Broglie wavelengths, $\delta _{\alpha \beta }=\left( \hbar ^{2}/\mu
_{\alpha \beta }k_{B}T\right) ^{1/2}$, where $\mu _{\alpha \beta }$ is the
reduced mass of the particles $\alpha $ and $\beta $. For large distances $%
r>\delta _{\alpha \beta }$ the potential becomes Coulomb, while for $%
r<\delta _{\alpha \beta }$ the Coulomb singularity is removed and $u_{\alpha
\beta }(0)=1/\delta _{\alpha \beta }$. By this the short range effects based
on the uncertainty principle are included \cite{mor59,kel63,deu81}. We point
out that the dependence on the physical parameters density and temperature
is implicitly contained in the parameters $\Gamma _{\alpha \beta }$ and $%
\delta _{\alpha \beta }$.

\section{Generalization of the Baranger-Mozer formulation to the TCP}
\label{sec:2}

\subsection{Preliminaries}
\label{sec:2.1}

The electric microfield distribution (MFD) $Q\left( \mathbf{E}\right) $ is
defined as the probability density of finding a field $\mathbf{E}$ at a
charge $Z_{R}e$, located at $\mathbf{r}_{0}$, in a TCP with $N_{i}$ ions and 
$N_{e}$ electrons. This system is described by classical statistical
mechanics in a canonical ensemble of $\left( N_{i}+N_{e}+1\right) $
particles, and temperature $T$. The normalized probability density of the
microfield $\mathbf{E}$ is then given by \cite{yan86,ner05} 
\begin{equation}
Q\left( \mathbf{E}\right) =\frac{1}{W}\int_{\Omega }e^{-\beta _{T}U\left( 
\mathcal{T}_{e},\mathcal{T}_{i},\mathbf{r}_{0}\right) }\delta \left( \mathbf{%
E}-\mathbf{E}\left( \mathcal{T}_{e},\mathcal{T}_{i},\mathbf{r}_{0}\right)
\right) d\mathbf{r}_{0}d\mathcal{T}_{e}d\mathcal{T}_{i},  \label{eq:4}
\end{equation}%
where $\beta _{T}=1/k_{B}T$, and $\mathcal{T}_{e}=\left\{ \mathbf{r}_{1},%
\mathbf{r}_{2}...\mathbf{r}_{N_{e}}\right\} $, $\mathcal{T}_{i}=\left\{ 
\mathbf{R}_{1},\mathbf{R}_{2}...\mathbf{R}_{N_{i}}\right\} $ are the
coordinates of electrons and ions, respectively, and 
\begin{equation}
W=\int_{\Omega }e^{-\beta _{T}U\left( \mathcal{T}_{e},\mathcal{T}_{i},%
\mathbf{r}_{0}\right) }d\mathbf{r}_{0}d\mathcal{T}_{e}d\mathcal{T}_{i}
\label{eq:5}
\end{equation}%
is the canonical partition function. $U\left( \mathcal{T}_{e},\mathcal{T}%
_{i},\mathbf{r}_{0}\right) $ is the potential energy of the configuration
given here by 
\begin{equation}
U\left( \mathcal{T}_{e},\mathcal{T}_{i},\mathbf{r}_{0}\right) =e_{S}^{2}%
\left[ \frac{1}{2}\sum_{\alpha ,\beta ,a,b}q_{\alpha }q_{\beta }u_{\alpha
\beta }(|\mathbf{r}_{a}^{\left( \alpha \right) }-\mathbf{r}_{b}^{\left(
\beta \right) }|)+Z_{R}\sum_{\alpha ,a}q_{\alpha }u_{\alpha R}(|\mathbf{r}%
_{0}-\mathbf{r}_{a}^{\left( \alpha \right) }|)\right]  \label{eq:6}
\end{equation}%
in terms of the pair interaction potentials $u_{\alpha \beta }\left(
r\right) $ and $u_{\alpha R}\left( r\right) $, where $\alpha ,\beta =e,i$, $%
\mathbf{r}_{a}^{\left( e\right) }=\mathbf{r}_{a}$, $\mathbf{r}_{a}^{\left(
i\right) }=\mathbf{R}_{a}$. In Eq.~(\ref{eq:6}) the first sum is restricted to $%
a\neq b$ for identical particles, $\alpha =\beta $. The total electrical
field $\mathbf{E}\left( \mathcal{T}_{e},\mathcal{T}_{i},\mathbf{r}%
_{0}\right) $ acting on the radiator is given by the superposition of
electronic and ionic single-particle fields 
\begin{equation}
\mathbf{E}\left( \mathcal{T}_{e},\mathcal{T}_{i},\mathbf{r}_{0}\right) =-%
\frac{1}{Z_{R}e}\mathbf{\nabla }_{0}U=\sum_{a=1}^{N_{e}}\mathbf{E}_{e}(%
\mathbf{r}_{0}-\mathbf{r}_{a})+\sum_{a=1}^{N_{i}}\mathbf{E}_{i}(\mathbf{r}%
_{0}-\mathbf{R}_{a}).  \label{eq:9}
\end{equation}%
As $\mathbf{E}_{\alpha }\left( \mathbf{r}\right) =\frac{\mathbf{r}}{r}%
E_{\alpha }\left( r\right) $, we obtain for the electronic and ionic
single-particle fields $E_{e}\left( r\right) =e_{F}u_{eR}^{\prime }\left(
r\right) $, $E_{i}\left( r\right) =-Ze_{F}u_{iR}^{\prime }\left( r\right) $,
where the prime indicates derivative with respect to $r$.

The spherical symmetric interaction between plasma particles and the isotropy
of the system allows to introduce the normalized microfield distribution
$P(E) =4\pi E^2 Q(E)$. It can be reexpressed in terms of the Fourier transform
of $Q(E)$ through 
\begin{equation}
P\left( E\right) =\frac{2E^{2}}{\pi }\int_{0}^{\infty }T\left( \kappa
\right) j_{0}\left( \kappa E\right) \kappa ^{2}d\kappa ,\quad T\left( \kappa
\right) =\int_{0}^{\infty }P\left( E\right) j_{0}\left( \kappa E\right) dE,
\label{eq:12}
\end{equation}%
and 
\begin{equation}
T\left( \boldsymbol{\kappa }\right) =\left\langle e^{i\boldsymbol{\kappa }%
\cdot \mathbf{E}}\right\rangle =\frac{1}{W}\int_{\Omega }\exp \left[ i%
\boldsymbol{\kappa }\cdot \mathbf{E}\left( \mathcal{T}_{e},\mathcal{T}_{i},%
\mathbf{r}_{0}\right) \right] e^{-\beta _{T}U\left( \mathcal{T}_{e},\mathcal{%
T}_{i},\mathbf{r}_{0}\right) }d\mathbf{r}_{0}d\mathcal{T}_{e}d\mathcal{T}%
_{i},  \label{eq:11}
\end{equation}%
where $\langle ...\rangle $ denotes a statistical average and $j_{0}(x)=\sin
x/x$ is the spherical Bessel function of order zero. The coefficients of the
expansion of the function $T(\kappa )$ at $\kappa \rightarrow 0$ yield the
even moments of the microfield distribution, 
\begin{equation}
T\left( \kappa \right) =1-\frac{\kappa ^{2}}{6}\left\langle
E^{2}\right\rangle +\frac{\kappa ^{4}}{120}\left\langle E^{4}\right\rangle
-...  \label{eq:13}
\end{equation}%
The similar expansion for the function $\mathcal{L}\left( \kappa \right) $
defined by $T(\kappa )=e^{-\mathcal{L}(\kappa )}$ yields 
\begin{equation}
\mathcal{L}\left( \kappa \right) =\frac{\kappa ^{2}}{6}\left\langle
E^{2}\right\rangle +\frac{\kappa ^{4}}{72}\left[ \left\langle
E^{2}\right\rangle ^{2}-\frac{3}{5}\left\langle E^{4}\right\rangle \right]
+...  \label{eq:14}
\end{equation}%
Therefore the Fourier transform of the MFD can be interpreted as a
generating function for microfield even moments. Equations (\ref{eq:4})-(\ref%
{eq:14}) describe the total MFD at the position $\mathbf{r}_{0}$ of the
radiator generated by both the statistically distributed ions and electrons
of the TCP. Since we are interested to calculate the MFD, Eq.~(\ref{eq:12}),
in an infinite system the statistical average of any quantity becomes
translationally invariant with respect to $\mathbf{r}_{0}$, and the location
of the test charge may be taken as the origin without loss of generality.

\subsection{Baranger-Mozer cluster expansion for the TCP}
\label{sec:2.2}

In this section we generalize the Baranger-Mozer (BM) cluster expansion
technique originally developed for one-component plasmas (see, e.g., \cite%
{bar59,moz60,hoo66,hoo68}) to the classical TCPs. This method results from
two transformations of Eq.~(\ref{eq:4}). The first is motivated by the fact
that the required average in Eq.~(\ref{eq:11}) is the product of electronic
and ionic single-particle functions, $e^{i\boldsymbol{\kappa }\cdot \mathbf{E%
}_{e}\left( \mathbf{r}_{a}\right) }$ and $e^{i\boldsymbol{\kappa }\cdot 
\mathbf{E}_{i}\left( \mathbf{R}_{a}\right) }$, which have a value close to
one over some volume of the system. This suggest a first transformation to
the set of single-particle functions 
\begin{equation}
\chi _{a}^{(\alpha )}\left( \boldsymbol{\kappa }\right) =e^{i\boldsymbol{%
\kappa }\cdot \mathbf{E}_{\alpha }\left( \mathbf{r}_{a}^{(\alpha )}\right)
}-1.  \label{eq:15}
\end{equation}%
Using this transformation the exponential factor in Eq.~(\ref{eq:11})
becomes (for simplicity we drop the coordinate $\mathbf{r}_{0}$ of the
radiator) 
\begin{equation}
\exp \left[ i\boldsymbol{\kappa }\cdot \mathbf{E}\left( \mathcal{T}_{e},%
\mathcal{T}_{i}\right) \right] =\prod_{a=1}^{N_{e}}\left[ 1+\chi
_{a}^{(e)}\left( \boldsymbol{\kappa }\right) \right] \prod_{b=1}^{N_{i}}%
\left[ 1+\chi _{b}^{(i)}\left( \boldsymbol{\kappa }\right) \right] .
\label{eq:16}
\end{equation}%
The introduced single-particle functions $\chi _{a}^{(\alpha )}$ have the
advantage to be zero over some volume. The spirit of this transformation to
the functions $\chi _{a}^{(\alpha )}$ is similar to the use of Mayer's $f$%
-functions for thermodynamic properties of gases \cite{may40}. Substituting
Eq.~(\ref{eq:16}) into Eq.~(\ref{eq:11}) leads in the thermodynamic limit
directly to the series 
\begin{eqnarray}
T\left( \boldsymbol{\kappa }\right) &=&1+\sum_{\alpha }\sum_{a=1}^{\infty }%
\frac{n_{\alpha }^{a}}{a!}\int \chi _{1}^{(\alpha )}\left( \boldsymbol{%
\kappa }\right) \chi _{2}^{(\alpha )}\left( \boldsymbol{\kappa }\right)
...\chi _{a}^{(\alpha )}\left( \boldsymbol{\kappa }\right) \mathcal{G}%
_{a}^{(\alpha )}(\mathcal{T}_{a}^{(\alpha )})d\mathcal{T}_{a}^{(\alpha )}
\label{eq:17} \\
&&+\sum_{a=1}^{\infty }\frac{n_{e}^{a}}{a!}\sum_{b=1}^{\infty }\frac{%
n_{i}^{b}}{b!}\int \chi _{1}^{(e)}\left( \boldsymbol{\kappa }\right) ...\chi
_{a}^{(e)}\left( \boldsymbol{\kappa }\right) \chi _{1}^{(i)}\left( 
\boldsymbol{\kappa }\right) ...\chi _{b}^{(i)}\left( \boldsymbol{\kappa }%
\right) \mathcal{G}_{ab}^{(ei)}(\mathcal{T}_{a}^{(e)},\mathcal{T}_{b}^{(i)})d%
\mathcal{T}_{a}^{(e)}d\mathcal{T}_{b}^{(i)},  \notag
\end{eqnarray}%
where $\mathcal{T}_{a}^{(\alpha )}=\{\mathbf{r}_{1}^{(\alpha )},\mathbf{r}%
_{2}^{(\alpha )},...\mathbf{r}_{a}^{(\alpha )}\}$ and $d\mathcal{T}%
_{a}^{(\alpha )}=d\mathbf{r}_{1}^{(\alpha )}d\mathbf{r}_{2}^{(\alpha )}...d%
\mathbf{r}_{a}^{(\alpha )}$. Here $\mathcal{G}_{a}^{(\alpha )}(\mathcal{T}%
_{a}^{(\alpha )})$ and $\mathcal{G}_{ab}^{(ei)}(\mathcal{T}_{a}^{(e)},%
\mathcal{T}_{b}^{(i)})$\ are equilibrium correlation functions. The first
quantity represents the probability density for $a$\ particles from plasma
species $\alpha $ at $\mathbf{r}_{1}^{(\alpha )}$, $\mathbf{r}_{2}^{(\alpha
)}$,$...\mathbf{r}_{a}^{(\alpha )}$, and the test particle at the origin.
The second one describes the correlations between $a$ electrons at $\mathbf{r%
}_{1}$, $\mathbf{r}_{2}$,$...\mathbf{r}_{a}$, and $b$ ions at $\mathbf{R}%
_{1} $, $\mathbf{R}_{2}$,$...\mathbf{R}_{b}$ involving the test particle.
The range of integration for each term in Eq.~(\ref{eq:17}) is now
restricted by the $\chi _{a}^{(\alpha )}\left( \boldsymbol{\kappa }\right) $%
-functions. However, this restriction is not uniform with respect to $%
\boldsymbol{\kappa }$, and particularly for large values of $\boldsymbol{%
\kappa }$ the functions $\chi _{a}^{(\alpha )}\left( \boldsymbol{\kappa }%
\right) $ can differ from zero over a correspondingly large volume.
Consequently, a second transformation is desirable, $T(\boldsymbol{\kappa }%
)=e^{-\mathcal{L}(\boldsymbol{\kappa })}$, where by a standard theorem of
equilibrium statistical mechanics (see, e.g., \cite{mun69}), $\mathcal{L}(%
\boldsymbol{\kappa })$ is determined from Eq.~(\ref{eq:17}) as 
\begin{equation}
\mathcal{L}\left( \boldsymbol{\kappa }\right) =-\sum_{\alpha
}\sum_{a=1}^{\infty }\frac{n_{\alpha }^{a}}{a!}h_{a}^{(\alpha )}\left( 
\boldsymbol{\kappa }\right) -\sum_{a=1}^{\infty }\frac{n_{e}^{a}}{a!}%
\sum_{b=1}^{\infty }\frac{n_{i}^{b}}{b!}h_{ab}^{(ei)}\left( \boldsymbol{%
\kappa }\right) .  \label{eq:18}
\end{equation}%
Here $h_{a}^{(\alpha )}\left( \boldsymbol{\kappa }\right) $ and $%
h_{ab}^{(ei)}\left( \boldsymbol{\kappa }\right) $ are given by 
\begin{equation}
h_{a}^{(\alpha )}\left( \boldsymbol{\kappa }\right) =\int \chi _{1}^{(\alpha
)}\left( \boldsymbol{\kappa }\right) \chi _{2}^{(\alpha )}\left( \boldsymbol{%
\kappa }\right) ...\chi _{a}^{(\alpha )}\left( \boldsymbol{\kappa }\right)
\ell _{a}^{(\alpha )}(\mathcal{T}_{a}^{(\alpha )})d\mathcal{T}_{a}^{(\alpha
)},  \label{eq:19}
\end{equation}%
\begin{equation}
h_{ab}^{(ei)}\left( \boldsymbol{\kappa }\right) =\int \chi _{1}^{(e)}\left( 
\boldsymbol{\kappa }\right) ...\chi _{a}^{(e)}\left( \boldsymbol{\kappa }%
\right) \chi _{1}^{(i)}\left( \boldsymbol{\kappa }\right) ...\chi
_{b}^{(i)}\left( \boldsymbol{\kappa }\right) \ell _{ab}^{(ei)}(\mathcal{T}%
_{a}^{(e)},\mathcal{T}_{b}^{(i)})d\mathcal{T}_{a}^{(e)}d\mathcal{T}%
_{b}^{(i)},  \label{eq:20}
\end{equation}%
and $\ell _{a}^{(\alpha )}$\ and $\ell _{ab}^{(ei)}$\ are the Ursell cluster
functions for TCP associated with the set of usual correlation functions $%
\mathcal{G}_{a}^{(\alpha )}$ and $\mathcal{G}_{ab}^{(ei)}$, respectively.
The functions $\ell _{a}^{(\alpha )}$\ for electron-electron and ion-ion
interactions are expressed by the correlation functions $\mathcal{G}%
_{a}^{(\alpha )}$\ in the same manner as in the case of corresponding OCP
(see, e.g., \cite{bar59,moz60,hoo66}). Therefore, as an example we consider
here explicitly only the functions $\ell _{ab}^{(ei)}$ which involve
electron-ion interactions, 
\begin{equation}
\ell _{11}^{(ei)}\left( \mathbf{r}_{1},\mathbf{R}_{1}\right) =\mathcal{G}%
_{11}^{(ei)}\left( \mathbf{r}_{1},\mathbf{R}_{1}\right) -\mathcal{G}%
_{1}^{(e)}\left( \mathbf{r}_{1}\right) \mathcal{G}_{1}^{(i)}\left( \mathbf{R}%
_{1}\right) ,  \label{eq:21}
\end{equation}%
\begin{eqnarray}
\ell _{12}^{(ei)}\left( \mathbf{r}_{1},\mathbf{R}_{1},\mathbf{R}_{2}\right)
&=&\mathcal{G}_{12}^{(ei)}\left( \mathbf{r}_{1},\mathbf{R}_{1},\mathbf{R}%
_{2}\right) -\mathcal{G}_{1}^{(e)}\left( \mathbf{r}_{1}\right) \mathcal{G}%
_{2}^{(i)}\left( \mathbf{R}_{1},\mathbf{R}_{2}\right) -\mathcal{G}%
_{1}^{(i)}\left( \mathbf{R}_{1}\right) \mathcal{G}_{11}^{(ei)}\left( \mathbf{%
r}_{1},\mathbf{R}_{2}\right)  \label{eq:22} \\
&&-\mathcal{G}_{1}^{(i)}\left( \mathbf{R}_{2}\right) \mathcal{G}%
_{11}^{(ei)}\left( \mathbf{r}_{1},\mathbf{R}_{1}\right) +2\mathcal{G}%
_{1}^{(e)}\left( \mathbf{r}_{1}\right) \mathcal{G}_{1}^{(i)}\left( \mathbf{R}%
_{1}\right) \mathcal{G}_{1}^{(i)}\left( \mathbf{R}_{2}\right) ,  \notag
\end{eqnarray}%
\begin{eqnarray}
\ell _{21}^{(ei)}\left( \mathbf{r}_{1},\mathbf{r}_{2},\mathbf{R}_{1}\right)
&=&\mathcal{G}_{21}^{(ei)}\left( \mathbf{r}_{1},\mathbf{r}_{2},\mathbf{R}%
_{1}\right) -\mathcal{G}_{1}^{(i)}\left( \mathbf{R}_{1}\right) \mathcal{G}%
_{2}^{(e)}\left( \mathbf{r}_{1},\mathbf{r}_{2}\right) -\mathcal{G}%
_{1}^{(e)}\left( \mathbf{r}_{1}\right) \mathcal{G}_{11}^{(ei)}\left( \mathbf{%
r}_{2},\mathbf{R}_{1}\right)  \label{eq:23} \\
&&-\mathcal{G}_{1}^{(e)}\left( \mathbf{r}_{2}\right) \mathcal{G}%
_{11}^{(ei)}\left( \mathbf{r}_{1},\mathbf{R}_{1}\right) +2\mathcal{G}%
_{1}^{(e)}\left( \mathbf{r}_{1}\right) \mathcal{G}_{1}^{(e)}\left( \mathbf{r}%
_{2}\right) \mathcal{G}_{1}^{(i)}\left( \mathbf{R}_{1}\right)  \notag
\end{eqnarray}%
etc., for pair and three-body correlations, respectively. The higher order
Ursell functions involving $\geqslant 4$ particles can be constructed
similarly. All single-particle Ursell functions are here expressed by the
ordinary pair correlation function between radiator and plasma particles of
species $\alpha $, $\ell _{1}^{(\alpha )}\left( \mathbf{r}\right) =\mathcal{G%
}_{1}^{(\alpha )}\left( \mathbf{r}\right) =g_{\alpha R}\left( r\right) $ 
\cite{ner05}. The significant difference between $\mathcal{G}$ and $\ell $
correlation functions is that the cluster functions $\ell $ vanish when any
members of the particles are sufficiently far apart, whereas the ordinary
correlation functions $\mathcal{G}$ have not this property. In particular,
for completely uncorrelated system of particles the many-body functions $%
\mathcal{G}$ tend to unity while $\ell \rightarrow 0$ beginning with
$\ell_{11}^{(ei)}$ and $\ell_{2}^{(\alpha)}$.
Hence, the range of the integrals in Eq.~(\ref{eq:18}%
) is controlled by both the functions $\chi _{a}^{(\alpha )}\left( 
\boldsymbol{\kappa }\right) $\ and the Ursell functions. The later restricts
the integration to volumes characterized by the correlation length which
depends on the thermodynamic-state condition but is independent of $%
\boldsymbol{\kappa }$. Qualitatively, therefore, the BM formalism provides a
series representation whose terms are controlled by the range of $\chi
_{a}^{(\alpha )}\left( \boldsymbol{\kappa }\right) $\ for small $\boldsymbol{%
\kappa }$\ and by the range of the Ursell functions for large $\boldsymbol{%
\kappa }$.

Equations~(\ref{eq:18})-(\ref{eq:20}) together with the relation $T(%
\boldsymbol{\kappa })=e^{-\mathcal{L}(\boldsymbol{\kappa })}$ constitutes
the generalization of the Baranger-Mozer cluster expansion technique to the
TCPs. In contrast to the BM theory developed for OCP Eq.~(\ref{eq:18}) now
involves terms which are responsible for electron-electron, ion-ion and
electron-ion attractive interactions. The BM result is easily recovered from
Eq.~(\ref{eq:18}) by neglecting the last term as well as one of two terms
responsible for the interactions between identical particles. In addition,
for ideal TCPs (Holtsmark limit) the second sum in Eq.~(\ref{eq:18}) and all
terms with $a\geqslant 2$ in the first sum vanish, and only the term with $%
a=1$ in the first sum contributes to the MFD. It is easy to see that this
term coincides with the result of Ref.~\cite{ner05} derived for ideal TCP.

From a practical point of view, it is too difficult to calculate correlation
functions of higher order than $\mathcal{G}_{2}^{(\alpha )}$ and $\mathcal{G}%
_{11}^{(ei)}$. For weakly coupled plasmas, the Ursell functions of order $%
\lambda +1$ are typically of the order of the plasma parameters to the power 
$\lambda $. Therefore Eq.~(\ref{eq:18}) is usually truncated at first,
second or in some cases at third order. If the correlation functions are
expanded as well with respect to the plasma parameters and truncated at the
corresponding order the MFD with Eqs.~(\ref{eq:18})-(\ref{eq:20}) can be
evaluated analytically (see, e.g., Refs.~\cite%
{bar59,moz60,hoo66,hoo68,hel81,dav04}). Obviously, such a method may be
successful only for weakly coupled plasmas. A possible extension to the
strongly coupled regimes requires that the BM representation sufficiently
rapidly converges to allow for a truncation after the first two terms. This
suggests, for instance, that the BM series could be improved if the bare
single-particle electric fields in the $\chi $-functions are replaced by
screened fields representing the effects of strong correlations in a plasma.
The formal procedure for carrying out such a "renormalization" has been
developed in Ref.~\cite{dom64}\ and has been previously employed for OCPs 
\cite{duf85}. In the next Sec.~\ref{sec:3} we extend this renormalization
procedure to the two-component plasmas.

\section{Renormalization of the Baranger-Mozer cluster series}
\label{sec:3}

\subsection{PMFEX as independent-particle model}
\label{sec:3.1}

In Refs.~\cite{ner05,ner06} we introduced the potential of mean force
exponential approximation (PMFEX) which links the MFD to the pair
correlation functions. To make further progress we briefly outline here the
basic concepts of the PMFEX approximation. We show that similarly to the
APEX (see, e.g., \cite{igl83,ala84,duf85}) PMFEX is also an effective
independent-particle model. Based on this assumption the MFD within PMFEX is
derived quite simply. If all interactions between the plasma particles,
except those with the radiator, are neglected, then the Ursell functions $%
\ell _{a}^{(\alpha )}$\ vanish for $a\geqslant 2$\ and\ $\ell _{ab}^{(ei)}(%
\mathcal{T}_{a}^{(e)},\mathcal{T}_{b}^{(i)})=0$ for arbitrary $a$ and $b$.
The BM series (\ref{eq:18}) then reduces to only the leading term, 
\begin{equation}
\mathcal{L}^{\left( 0\right) }\left( \boldsymbol{\kappa }\right)
=-\sum_{\alpha }n_{\alpha }\int \chi _{1}^{(\alpha )}\left( \boldsymbol{%
\kappa }\right) g_{\alpha R}^{\left( 0\right) }\left( \mathbf{r}_{1}\right) d%
\mathbf{r}_{1}.  \label{eq:24}
\end{equation}%
The superscript $(0)$\ denotes the corresponding quantity without
interactions among plasma particles. The PMFEX approach retains the
independent-particle form of (\ref{eq:24}), 
\begin{equation}
\mathcal{L}^{\left( 0\right) }\left( \boldsymbol{\kappa }\right) \rightarrow 
\mathcal{L}\left( \boldsymbol{\kappa }\right) =-\sum_{\alpha }n_{\alpha
}\int \psi _{1}^{(\alpha )}\left( \boldsymbol{\kappa }\right) g_{\alpha
R}^{\ast }\left( \mathbf{r}_{1}\right) d\mathbf{r}_{1},  \label{eq:25}
\end{equation}%
with the assumption that the important effects of correlations can be
accounted for by an effective pair distribution function $g_{\alpha R}^{\ast
}\left( \mathbf{r}_{1}\right) $ and a screened field $\boldsymbol{\mathcal{E}%
}_{\alpha }\left( \mathbf{r}_{1}\right) $\ replacing the single-particle
field $\mathbf{E}_{\alpha }\left( \mathbf{r}_{1}\right) $\ in $\chi
_{1}^{(\alpha )}\left( \boldsymbol{\kappa }\right) $, 
\begin{equation}
\psi _{1}^{(\alpha )}\left( \boldsymbol{\kappa }\right) =e^{i\boldsymbol{%
\kappa }\cdot \boldsymbol{\mathcal{E}}_{\alpha }\left( \mathbf{r}_{1}\right)
}-1.  \label{eq:26}
\end{equation}%
Some constraint must be imposed to determine $g_{\alpha R}^{\ast }\left( 
\mathbf{r}_{1}\right) $. As in the case of APEX \cite{igl83} this is a
requirement that the effective "quasiparticle" field due to the effective
charge density at $\mathbf{r}_{1}$ is equal to the corresponding exact field 
\begin{equation}
g_{\alpha R}^{\ast }\left( \mathbf{r}_{1}\right) \boldsymbol{\mathcal{E}}%
_{\alpha }\left( \mathbf{r}_{1}\right) =g_{\alpha R}\left( \mathbf{r}%
_{1}\right) \mathbf{E}_{\alpha }\left( \mathbf{r}_{1}\right) .  \label{eq:27}
\end{equation}%
Assuming the spherical symmetric interactions with\ $\boldsymbol{\mathcal{E}}%
_{\alpha }\left( \mathbf{r}\right) =\frac{\mathbf{r}}{r}\mathcal{E}_{\alpha
}\left( r\right) $\ and $\mathbf{E}_{\alpha }\left( \mathbf{r}\right) =\frac{%
\mathbf{r}}{r}E_{\alpha }\left( r\right) $,\ this condition then define the
PMFEX model as 
\begin{equation}
\mathcal{L}_{\mathrm{PMFEX}}\left( \kappa \right) =\sum_{\alpha }4\pi
n_{\alpha }\int_{0}^{\infty }E_{\alpha }\left( r\right) \frac{1-j_{0}\left(
\kappa \mathcal{E}_{\alpha }\left( r\right) \right) }{\mathcal{E}_{\alpha
}\left( r\right) }g_{\alpha R}\left( r\right) r^{2}dr.  \label{eq:28}
\end{equation}%
In practice $g_{\alpha R}\left( r\right) $\ is calculated from the
hypernetted-chain (HNC) integral equation \cite{han76,bau80} extended to the
TCP (see, e.g., Refs.~\cite{ner05,ner06}).

In the APEX model originally developed for the classical ionic OCP with bare
Coulomb interaction to determine the effective field a second requirement is
needed. In Ref.~\cite{igl83} the effective field $\mathcal{E}(r)$ is assumed
as a Debye-H\"{u}ckel-type screened interaction with unknown screening
length. This free parameter is then adjusted in such a way to satisfy the
exact second moment requirement. In contrast to the APEX the effective
fields $\mathcal{E}_{\alpha }\left( r\right) $ in the PMFEX model are
obtained automatically (see \cite{ner05} for details) employing the
thermodynamic perturbation theory \cite{zwa54}. We recall here these
effective fields for completeness. For simplicity assuming Coulomb potential
for the repulsive interactions, i.e. $\delta _{ee}=\delta _{ii}=0$
(electron-electron and ion-ion interactions) and regularized one (see Eq.~(%
\ref{eq:3})) for electron-ion interactions these fields read \cite{ner05} 
\begin{equation}
\mathcal{E}_{\alpha }\left( r\right) =E_{\alpha }\left( r\right) \left\{ 1+%
\frac{4\pi n_{\alpha }}{g_{\alpha R}\left( r\right) }\int_{0}^{r}\left[
g_{\alpha \alpha }\left( \rho \right) -g_{ei}\left( \rho \right) \right]
\rho ^{2}d\rho \right\} ,  \label{eq:29}
\end{equation}%
where $E_{\alpha }\left( r\right) =q_{\alpha }e_{F}/r^{2}$ are the bare
single-particle fields. Based on the derivations given in Ref.~\cite{ner05}
it is straightforward to show that the effective fields $\mathcal{E}_{\alpha
}\left( r\right) $ can be alternatively represented as the logarithmic
derivative of the radial distribution functions (RDF) $g_{\alpha R}\left(
r\right) $ as 
\begin{equation}
\mathcal{E}_{\alpha }\left( r\right) =\frac{k_{B}T}{Z_{R}e}\frac{\partial }{%
\partial r}\left[ \ln g_{\alpha R}\left( r\right) \right]  \label{eq:30}
\end{equation}%
which is known as the potential of mean force (PMF) approximation proposed
by Yan and Ichimaru \cite{yan86} (see also \cite{han76}). Using Eq.~(\ref%
{eq:14}) we expand (\ref{eq:28}) around a value $\kappa =0$\ and find the
second moment of the MFD in the PMFEX model 
\begin{equation}
\left\langle E^{2}\right\rangle _{\mathrm{PMFEX}}=\sum_{\alpha }4\pi
n_{\alpha }\int_{0}^{\infty }\mathcal{E}_{\alpha }\left( r\right) E_{\alpha
}\left( r\right) g_{\alpha R}\left( r\right) r^{2}dr.  \label{eq:30a}
\end{equation}%
Then, introducing Eq.~(\ref{eq:30}) in Eq.~(\ref{eq:30a}) automatically
satisfies the exact sum rule, $\langle E^{2}\rangle _{\mathrm{PMFEX}%
}=\langle E^{2}\rangle $, without any adjustable parameter. Here $%
\left\langle E^{2}\right\rangle $\ is the exact second moment of the MFD
derived in Ref.~\cite{ner05} for the TCP, see Eq.~(\ref{eq:app1}). Thus, if
the $g_{\alpha R}(r)$ are known the MFD within the PMFEX model and with the
exact second moment can be calculated using Eqs.~(\ref{eq:12}), (\ref{eq:28}),
and (\ref{eq:30}).

The PMFEX effective field, Eq.~(\ref{eq:30}), is valid for charged radiator
with $Z_{R}\neq 0$. For a neutral radiator, i.e., in the limit $%
Z_{R}\rightarrow 0$, the PMF ansatz (\ref{eq:30}) is not applicable, as in
this case the RDFs tend to unity, $g_{\alpha R}\rightarrow 1$, and $\ln
g_{\alpha R}\left( r\right) \rightarrow 0$. Based on the derivations given
in Refs.~\cite{ner05,ner06} (see also Appendix~\ref{sec:app1} for details)
the correct limit $Z_{R}\rightarrow 0$ can, however, be done and results in
the effective field of Eq.~(\ref{eq:29}), by setting $g_{\alpha R}\left(
r\right) =1$.

\subsection{Renormalized cluster series}
\label{sec:3.2}

The comments at the end of Sec.~\ref{sec:2.2} and the success of PMFEX
indicate that the Baranger-Mozer series could be improved if the
single-particle fields in the functions $\chi _{a}^{(\alpha )}\left( 
\boldsymbol{\kappa }\right) $ is replaced by a screened fields representing
the effects of correlations in strongly coupled plasmas. In this spirit, a
new functional series is obtained in terms of the renormalized functions $%
\psi _{a}^{(\alpha )}\left( \boldsymbol{\kappa }\right) $\ of Eq.~(\ref%
{eq:26}) by the definition $\mathcal{L}^{\ast }\left[ \psi _{a};\boldsymbol{%
\kappa }\right] =\mathcal{L}\left[ \chi _{a};\boldsymbol{\kappa }\right] $\
which is easily obtained from the BM series Eqs.~(\ref{eq:18})-(\ref{eq:20})
and the functional relationship between $\chi _{a}^{(\alpha )}$ and $\psi
_{a}^{(\alpha )}$. Introducing the functions $\mathcal{R}_{\alpha }\left(
r\right) $ by the relation $\mathcal{E}_{\alpha }\left( r\right) =\mathcal{R}%
_{\alpha }\left( r\right) E_{\alpha }\left( r\right) $ (we assume spherical
symmetric interactions) with the effective $\mathcal{E}_{\alpha }\left(
r\right) $ and single-particle $E_{\alpha }\left( r\right) $ fields, the
functional relation between $\chi _{a}^{(\alpha )}$ and $\psi _{a}^{(\alpha
)}$ is established as 
\begin{equation}
\chi _{a}^{(\alpha )}\left( \boldsymbol{\kappa }\right) =\left[ 1+\psi
_{a}^{(\alpha )}\left( \boldsymbol{\kappa }\right) \right] ^{1/\mathcal{R}%
_{\alpha }\left( r_{a}\right) }-1\mathbf{.}  \label{eq:31}
\end{equation}%
To obtain the new function $\mathcal{L}^{\ast }\left[ \psi _{a};\boldsymbol{%
\kappa }\right] $\ which must be multi-linear with respect to $\psi _{a}$\
(cf. Eqs.~(\ref{eq:18})-(\ref{eq:20})) we expand the $\chi _{a}^{(\alpha )}$%
-functions in Eq.~(\ref{eq:31})\ with respect to $\psi _{a}^{(\alpha )}$%
-functions \cite{gra80} 
\begin{eqnarray}
\chi _{a}^{(\alpha )}\left( \boldsymbol{\kappa }\right) &=&\frac{\psi
_{a}^{(\alpha )}\left( \boldsymbol{\kappa }\right) }{\mathcal{R}_{\alpha
}\left( r_{a}\right) }+\frac{\left[ \psi _{a}^{(\alpha )}\left( \boldsymbol{%
\kappa }\right) \right] ^{2}}{2!}\frac{1}{\mathcal{R}_{\alpha }\left(
r_{a}\right) }\left[ \frac{1}{\mathcal{R}_{\alpha }\left( r_{a}\right) }-1%
\right]  \label{eq:32} \\
&&+\frac{\left[ \psi _{a}^{(\alpha )}\left( \boldsymbol{\kappa }\right) %
\right] ^{3}}{3!}\frac{1}{\mathcal{R}_{\alpha }\left( r_{a}\right) }\left[ 
\frac{1}{\mathcal{R}_{\alpha }\left( r_{a}\right) }-1\right] \left[ \frac{1}{%
\mathcal{R}_{\alpha }\left( r_{a}\right) }-2\right] +...  \notag
\end{eqnarray}%
Then elimination of $\chi _{a}^{(\alpha )}$\ on the right side of equation $%
\mathcal{L}^{\ast }\left[ \psi _{a};\boldsymbol{\kappa }\right] =\mathcal{L}%
\left[ \chi _{a};\boldsymbol{\kappa }\right] $\ using Eq.~(\ref{eq:32})
yields the desired renormalized cluster series, 
\begin{equation}
\mathcal{L}^{\ast }\left[ \psi _{a};\boldsymbol{\kappa }\right]
=-\sum_{\alpha }\sum_{a=1}^{\infty }\frac{n_{\alpha }^{a}}{a!}H_{a}^{(\alpha
)}\left( \boldsymbol{\kappa }\right) -\sum_{a=1}^{\infty }\frac{n_{e}^{a}}{a!%
}\sum_{b=1}^{\infty }\frac{n_{i}^{b}}{b!}H_{ab}^{(ei)}\left( \boldsymbol{%
\kappa }\right) .  \label{eq:33}
\end{equation}%
Here 
\begin{equation}
H_{a}^{(\alpha )}\left( \boldsymbol{\kappa }\right) =\int \psi _{1}^{(\alpha
)}\left( \boldsymbol{\kappa }\right) \psi _{2}^{(\alpha )}\left( \boldsymbol{%
\kappa }\right) ...\psi _{a}^{(\alpha )}\left( \boldsymbol{\kappa }\right)
L_{a}^{(\alpha )}(\mathcal{T}_{a}^{(\alpha )})d\mathcal{T}_{a}^{(\alpha )},
\label{eq:34}
\end{equation}%
\begin{eqnarray}
H_{ab}^{(ei)}\left( \boldsymbol{\kappa }\right) &=&\int \psi _{1}^{\left(
e\right) }\left( \boldsymbol{\kappa }\right) \psi _{2}^{\left( e\right)
}\left( \boldsymbol{\kappa }\right) ...\psi _{a}^{(e)}\left( \boldsymbol{%
\kappa }\right) \psi _{1}^{(i)}\left( \boldsymbol{\kappa }\right) \psi
_{2}^{(i)}\left( \boldsymbol{\kappa }\right) ...\psi _{b}^{(i)}\left( 
\boldsymbol{\kappa }\right)  \label{eq:35} \\
&&\times L_{ab}^{(ei)}(\mathcal{T}_{a}^{(e)},\mathcal{T}_{b}^{(i)})d\mathcal{%
T}_{a}^{(e)}d\mathcal{T}_{b}^{(i)}  \notag
\end{eqnarray}%
with the generalized Ursell functions $L_{a}^{(\alpha )}$ and $L_{ab}^{(ei)}$
which are recognized as the functional derivatives of $H_{a}^{(\alpha
)}\left( \boldsymbol{\kappa }\right) $ and $H_{ab}^{(ei)}\left( \boldsymbol{%
\kappa }\right) $, respectively. Explicitly the first and second order
generalized Ursell functions are found to be 
\begin{eqnarray}
L_{1}^{(\alpha )}\left( r_{1}\right) &=&\frac{g_{\alpha R}\left(
r_{1}\right) }{\mathcal{R}_{\alpha }\left( r_{1}\right) },  \label{eq:38} \\
L_{2}^{(\alpha )}\left( \mathbf{r}_{1},\mathbf{r}_{2}\right) &=&\frac{%
g_{\alpha \alpha }\left( \left\vert \mathbf{r}_{1}-\mathbf{r}_{2}\right\vert
\right) -g_{\alpha R}\left( r_{1}\right) g_{\alpha R}\left( r_{2}\right) }{%
\mathcal{R}_{\alpha }\left( r_{1}\right) \mathcal{R}_{\alpha }\left(
r_{2}\right) }-\frac{1}{n_{\alpha }}\delta \left( \mathbf{r}_{1}-\mathbf{r}%
_{2}\right) g_{\alpha R}\left( r_{1}\right) \frac{\mathcal{R}_{\alpha
}\left( r_{1}\right) -1}{\mathcal{R}_{\alpha }^{2}\left( r_{1}\right) },
\label{eq:39} \\
L_{11}^{(ei)}\left( \mathbf{r}_{1},\mathbf{R}_{1}\right) &=&\frac{%
g_{ei}\left( \left\vert \mathbf{r}_{1}-\mathbf{R}_{1}\right\vert \right)
-g_{eR}\left( r_{1}\right) g_{iR}\left( R_{1}\right) }{\mathcal{R}_{e}\left(
r_{1}\right) \mathcal{R}_{i}\left( R_{1}\right) }  \label{eq:40}
\end{eqnarray}%
which result the first three terms in Eq.~(\ref{eq:33}) 
\begin{eqnarray}
\mathcal{L}^{\ast }\left( \boldsymbol{\kappa }\right) &=&\mathcal{L}\left( 
\boldsymbol{\kappa }\right) =-\sum_{\alpha }n_{\alpha }\int \frac{\psi
_{1}^{(\alpha )}\left( \boldsymbol{\kappa }\right) }{\mathcal{R}_{\alpha
}\left( r\right) }g_{\alpha R}\left( r\right) d\mathbf{r}  \notag \\
&&-\sum_{\alpha }\frac{n_{\alpha }^{2}}{2}\left\{ \int \frac{\psi
_{1}^{(\alpha )}\left( \boldsymbol{\kappa }\right) \psi _{2}^{(\alpha
)}\left( \boldsymbol{\kappa }\right) }{\mathcal{R}_{\alpha }\left(
r_{1}\right) \mathcal{R}_{\alpha }\left( r_{2}\right) }\left[ g_{\alpha
\alpha }\left( \left\vert \mathbf{r}_{1}-\mathbf{r}_{2}\right\vert \right)
-g_{\alpha R}\left( r_{1}\right) g_{\alpha R}\left( r_{2}\right) \right] d%
\mathbf{r}_{1}d\mathbf{r}_{2}\right.  \label{eq:41} \\
&&\left. +\frac{1}{n_{\alpha }}\int \left[ \psi _{1}^{(\alpha )}\left( 
\boldsymbol{\kappa }\right) \right] ^{2}\frac{1-\mathcal{R}_{\alpha }\left(
r_{1}\right) }{\mathcal{R}_{\alpha }^{2}\left( r_{1}\right) }g_{\alpha
R}\left( r_{1}\right) d\mathbf{r}_{1}\right\}  \notag \\
&&-n_{e}n_{i}\int \frac{\psi _{1}^{(e)}\left( \boldsymbol{\kappa }\right)
\psi _{1}^{(i)}\left( \boldsymbol{\kappa }\right) }{\mathcal{R}_{e}\left(
r_{1}\right) \mathcal{R}_{i}\left( R_{1}\right) }\left[ g_{ei}\left(
\left\vert \mathbf{r}_{1}-\mathbf{R}_{1}\right\vert \right) -g_{eR}\left(
r_{1}\right) g_{iR}\left( R_{1}\right) \right] d\mathbf{r}_{1}d\mathbf{R}%
_{1}.  \notag
\end{eqnarray}%
The first term of Eq.~(\ref{eq:41}) is seen to be precisely PMFEX (cf. Eq.~(%
\ref{eq:25}) and (\ref{eq:27})). The factor $\mathcal{R}_{\alpha }\left(
r\right) $\ occurs automatically here from the renormalization and
eliminates the assumption (\ref{eq:27}) of PMFEX.

Finally the angular integration in Eq.~(\ref{eq:41}) can be performed using
spherical harmonic expansion \cite{gra80}. This yields 
\begin{equation}
\mathcal{L}\left( \kappa \right) =\mathcal{L}_{\mathrm{PMFEX}}\left( \kappa
\right) +\Delta \mathcal{L}\left( \kappa \right) ,  \label{eq:42}
\end{equation}%
where the first term is the PMFEX result, Eq.~(\ref{eq:28}), and the second
term represents the corrections due to the renormalization, 
\begin{eqnarray}
\Delta \mathcal{L}\left( \kappa \right) &=&2\pi \sum_{\alpha }n_{\alpha
}\int_{0}^{\infty }\left[ j_{0}\left( 2\kappa \mathcal{E}_{\alpha }\left(
r\right) \right) -2j_{0}\left( \kappa \mathcal{E}_{\alpha }\left( r\right)
\right) +1\right] \frac{\mathcal{R}_{\alpha }\left( r\right) -1}{\mathcal{R}%
_{\alpha }^{2}\left( r\right) }g_{\alpha R}\left( r\right) r^{2}dr
\label{eq:43} \\
&&-4\sum_{\alpha }n_{\alpha }\int_{0}^{\infty }G_{\alpha \alpha }\left(
\kappa ,k\right) \left[ S_{\alpha \alpha }\left( k\right) -1\right] k^{2}dk-8%
\frac{n_{e}n_{i}}{n}\int_{0}^{\infty }G_{ei}\left( \kappa ,k\right)
S_{ei}\left( k\right) k^{2}dk  \notag \\
&&+\frac{1}{2}\left[ \mathcal{L}_{\mathrm{PMFEX}}^{2}\left( \kappa \right) -%
\mathcal{L}_{0}^{2}\left( \kappa \right) \right] .  \notag
\end{eqnarray}%
Here $\mathcal{L}_{0}\left( \kappa \right) $ is the $\mathcal{L}_{\mathrm{%
PMFEX}}\left( \kappa \right) $ given by Eq.~(\ref{eq:28}) but with $%
g_{\alpha R}\left( r\right) =1$, 
\begin{equation}
\mathcal{L}_{0}\left( \kappa \right) =\sum_{\alpha }4\pi n_{\alpha
}\int_{0}^{\infty }\frac{1-j_{0}\left( \kappa \mathcal{E}_{\alpha }\left(
r\right) \right) }{\mathcal{R}_{\alpha }\left( r\right) }r^{2}dr.
\label{eq:44}
\end{equation}%
In Eq.~(\ref{eq:43}) the term $S_{\alpha \beta }\left( k\right) $ (with $%
\alpha ,\beta =e,i$) is the static structure factor for the two-component
plasma, 
\begin{equation}
S_{\alpha \beta }\left( k\right) =\delta _{\alpha \beta }+4\pi n_{\alpha
\beta }\int_{0}^{\infty }\left[ g_{\alpha \beta }\left( r\right) -1\right]
j_{0}\left( kr\right) r^{2}dr  \label{eq:45}
\end{equation}%
with $\delta _{\alpha \alpha }=1$, $\delta _{ei}=0$, $n_{\alpha \alpha
}=n_{\alpha }$, $n_{ei}=n=n_{e}+n_{i}$, and $G_{\alpha \beta }\left( \kappa
,k\right) $ is defined by 
\begin{equation}
G_{\alpha \beta }\left( \kappa ,k\right) =\sum_{l=0}^{\infty }\left(
-1\right) ^{l}\left( 2l+1\right) J_{l}^{\left( \alpha \right) }\left( \kappa
,k\right) J_{l}^{\left( \beta \right) }\left( \kappa ,k\right) ,
\label{eq:46}
\end{equation}%
\begin{equation}
J_{l}^{\left( \alpha \right) }\left( \kappa ,k\right) =\int_{0}^{\infty
}j_{l}\left( kr\right) \left[ j_{l}\left( \kappa \mathcal{E}_{\alpha
}(r)\right) -\delta _{l0}\right] \frac{r^{2}dr}{\mathcal{R}_{\alpha }\left(
r\right) }.  \label{eq:47}
\end{equation}%
Also, $j_{l}\left( x\right) $ is the spherical Bessel function of order $l$.
It should be noted that for deriving the renormalized series Eq.~(\ref{eq:33}%
) we do not use explicit functional form of $\mathcal{E}_{\alpha }(r)$.
Hence, in spite of its important role in the theory, the precise functional
form of the effective field $\mathcal{E}_{\alpha }(r)$\ may remain
arbitrary. Moreover, it is straightforward to show that the generating
function Eq.~(\ref{eq:18}) as well as its renormalized version Eq.~(\ref%
{eq:33}) yield the exact even moments of the MFD according to Eq.~(\ref%
{eq:14}) independently of the functional form of $\mathcal{E}_{\alpha }(r)$.
In the present context it might appear more reasonable to choose $\mathcal{E}%
_{\alpha }(r)$ to improve convergence of the renormalized series Eq.~(\ref%
{eq:33}). A similar procedure is used in thermodynamic perturbation theory
where the corresponding parameters of the leading term are chosen to make
the next order terms vanish \cite{ree73}. But this is not possible
without making $\mathcal{E}_{\alpha }(r)$\ a function of $\boldsymbol{\kappa 
}$ (see similar discussion in Ref.~\cite{duf85} in the context of APEX).
Here we consider for $\mathcal{E}_{\alpha }(r)$\ the mean-force field as
derived in Ref.~\cite{ner05} for the TCP (see also Eq.~(\ref{eq:30}) as well
as Eq.~(\ref{eq:29}) for the repulsive Coulomb interactions) and which
appears automatically from the second moment condition $\left. \partial ^{2}%
\mathcal{L}^{\ast }/\partial \kappa ^{2}\right\vert _{\kappa =0}=\frac{1}{3}%
\langle E^{2}\rangle $ for the renormalized function $\mathcal{L}^{\ast }$ (%
\ref{eq:33}). The theoretical scheme resulting from Eqs.~(\ref{eq:42})-(\ref%
{eq:47})\ is abbreviated as PMFEX+. It agrees for neutral points quite well
with the molecular dynamic (MD) simulation results, as we will show in the
next section. It is straightforward to show that the correction
$\Delta \mathcal{L}(\kappa ) $ in Eq.~(\ref{eq:42}) behaves as $\sim
\kappa ^{4}$ at $\kappa \rightarrow 0$ and, therefore, does not contribute
to the second moment. Then the quantity $\langle E^{2}\rangle $\ receives
contribution only from the PMFEX term in Eq.~(\ref{eq:42}).

\section{Application to the neutral radiator}
\label{sec:4}

In the case of neutral radiator, $Z_{R}=0$, the plasma-radiator correlation
function is $g_{\alpha R}\left( r\right) =1$ and Eq.~(\ref{eq:43}) is
simplified. In particular, the last term vanishes since $\mathcal{L}_{%
\mathrm{PMFEX}}\left( \kappa \right) =\mathcal{L}_{0}\left( \kappa \right) $%
\ for $g_{\alpha R}\left( r\right) =1$. Thus, the generating function $%
\mathcal{L}\left( \kappa \right) $ of the MFD in the PMFEX+ model is given
by $\mathcal{L}\left( \kappa \right) =\mathcal{L}_{0}\left( \kappa \right)
+\Delta \mathcal{L}\left( \kappa \right) $, where $\mathcal{L}_{0}\left(
\kappa \right) $ is the ordinary PMFEX expression, with $g_{\alpha R}\left(
r\right) =1$, see Eq.~(\ref{eq:44}). The higher order corrections are
involved in the second term, $\Delta \mathcal{L}\left( \kappa \right) $. 
The functions $\mathcal{R}_{\alpha }\left( r\right) $ are obtained from Eq.~(\ref%
{eq:29}), 
\begin{equation}
\mathcal{R}_{\alpha }\left( r\right) =1+4\pi n_{\alpha }\int_{0}^{r}\left[
g_{\alpha \alpha }(\rho )-g_{ei}(\rho )\right] \rho ^{2}d\rho  \label{eq:48}
\end{equation}%
which in the limit of small plasma-parameters ($\Gamma _{ee}\rightarrow 0$)
can be evaluated explicitly using the Debye-H\"{u}ckel approximation, $%
\mathcal{R}_{e}\left( r\right) =\mathcal{R}_{i}\left( r\right) \simeq \left( 
\frac{r}{\lambda }+1\right) e^{-r/\lambda }$. Here $\lambda ^{2}=\varepsilon
_{0}k_{B}T/(Zne^{2})$ is the Debye screening length. Thus similar to the
standard PMFEX, the PMFEX+ approximation links the MFD to the RDFs. To obtain
explicit results for the MFD the corresponding RDFs and the static structure
factors must be determined first. This is done by solving numerically the
hypernetted chain (HNC) integral equations \cite{han76,bau80} for the TCPs 
under consideration (see Ref.~\cite{ner05} and references therein for
more details). For simplicity we here assume bare Coulomb interaction for 
electron-electron and ion-ion, i.e. $\delta _{ee}\simeq 0$ and $\delta _{ii}\simeq 0$,
and a regularized ion-electron interaction with a parameter $\delta _{ei}=\delta $
fixed to $\delta =0.2a$ or $\delta =0.4a$, where $a=\left( 4\pi n/3\right) ^{-1/3}$ 
is the Wigner-Seitz radius. For this kind of TCPs, the HNC method has already been
extensively tested and evaluated by comparison of the resulting RDFs with those 
obtained by classical MD simulations \cite{ner05}. 
\begin{table}[tbp]
\caption{The critical values $\sigma _{c}(Z,\delta )$ for
some values of the ion charge $Z$ and three values of $\delta =0.1a$%
, $0.2a$ and $0.4a$.}
\label{tab:1}
\begin{center}
\begin{tabular}{p{1.1cm}p{1.1cm}p{1.1cm}p{1.1cm}p{1.1cm}p{1.1cm}p{1.1cm}}
\hline\hline
$\delta$ & H$^{+}$ & Li$^{3+}$ & B$^{5+}$ & N$^{7+}$ & Al$^{13+}$ & Ca$%
^{20+} $ \\ \hline
$0.1a$ & 6.87 & 7.59 & 7.67 & 7.62 & 7.59 & 7.63 \\ 
$0.2a$ & 6.55 & 7.68 & 7.52 & 7.13 & 6.66 & 6.48 \\ 
$0.4a$ & 8.90 & 12.73 & 10.76 & 9.39 & 6.75 & 6.00 \\ \hline
\end{tabular}%
\end{center}
\end{table}
For the numerical solution of the HNC scheme the dimensionless parameter 
$\sigma_{ei}=Ze_{S}^{2}u_{ei}\left( 0\right) /k_{B}T=\Gamma _{ei}\left( a/\delta
\right) $, i.e.~the maximum value of the electron-ion interaction energy 
in units of $k_{B}T$, plays an important role (see also \cite{tal02,ner05}). 
Within our numerical treatment of the HNC equations a parameter regime with 
$\sigma _{ei}<\sigma_{c}(Z,\delta )$ is accessible, where the critical values 
$\sigma _{c}$ for $\delta =0.1a$, $0.2a$ and $0.4a$ and the different studied TCPs 
(H$^{+}$, Li$^{3+}$, B$^{5+}$, N$^{7+}$, Al$^{13+}$ and Ca$^{20+}$) 
are given in Table~\ref{tab:1}. Beyond this value the HNC numerical procedure does 
either not converge or ends up in unphysical solutions. A similar behavior has been 
reported in Ref.~\cite{tal02} for the case of an ion embedded in electrons. 
With the RDFs and $S_{\alpha \beta }\left( k\right) $ provided by the HNC scheme the MFD $P(E)$
in the PMFEX+ model is then calculated via Eqs.~(\ref{eq:12}) and (\ref%
{eq:42})-(\ref{eq:48}) by standard numerical differentiation and integration
methods \cite{numrep}. In practice it appears sufficient to terminate the
sum in Eq.~(\ref{eq:46}) at $l=5$.

To test the PMFEX+ approximation and improvements compared to the standard PMFEX
approach the calculated MFDs are confronted with those obtained by classical
MD simulations. In the MD simulations the classical equations of
motion are numerically integrated for $N_{i}$ ions and $N_{e}=ZN_{i}$
electrons interacting via the regularized potentials $u_{\alpha \beta }(r)$.
Such MD simulations have already been extensively tested and successfully
applied for investigations of the dynamic properties 
\cite{psch01,sel01,psch03,zwi03,mor05,zwi06} and 
the MFDs \cite{ner05,ner06} of a TCP with regularized
potentials (see also Ref.~\cite{cal07} and references therein).
For our present considerations we investigated the specific cases of 
H$^{+}$, Li$^{3+}$, B$^{5+}$, N$^{7+}$, Al$^{13+}$ and Ca$^{20+}$ TCPs 
with symmetric (hydrogen) and strongly asymmetric densities of the 
plasma species and certain values of the coupling parameters 
$\Gamma _{\alpha \beta }$ with $\alpha ,\beta =e,i$. These 
coupling parameters $\Gamma _{\alpha \beta }$, at a given parameter $\delta$,
are chosen to avoid the mentioned numerical difficulties with the HNC scheme and 
the formation of unphysical bound states in the classical MD simulations.
This implies in particular that for highly charged
plasma ions with $Z\gg 1$ plasma states with strongly correlated ions ($\Gamma _{ii}%
\gg 1$) and strong electron-ion interactions ($\Gamma _{ei}\sim 1 $) are accessible,
while the parameter $\Gamma _{ee}$ has to remain quite small, $\Gamma_{ee}\simeq \Gamma%
_{ei}/Z\ll 1$. But in the case of highly charged ions even for $\Gamma _{ee}\ll 1$
the electrons may be strongly correlated due to nonlinear effects 
as will be discussed below.
The MD simulations providing the present MFDs and RDFs have been performed using
$N_{\mathrm{MD}}=N_{i}+N_{e} = 5376$ particles. Further details on the 
applied MD technique can be found in Refs.~\cite{psch01,sel01,psch03,zwi03}.
As the theoretical models and the numerical solutions depend directly on the
coupling parameters $\Gamma_{\alpha \beta }$ and the regularization
parameter $\delta $ we discuss, as e.g.~in Refs.~\cite{tal02,ner05,ner06}, 
our results in terms of these parameters
rather than in the underlying physical values of density and temperature.

\begin{figure}[tbp]
\includegraphics[width=8cm]{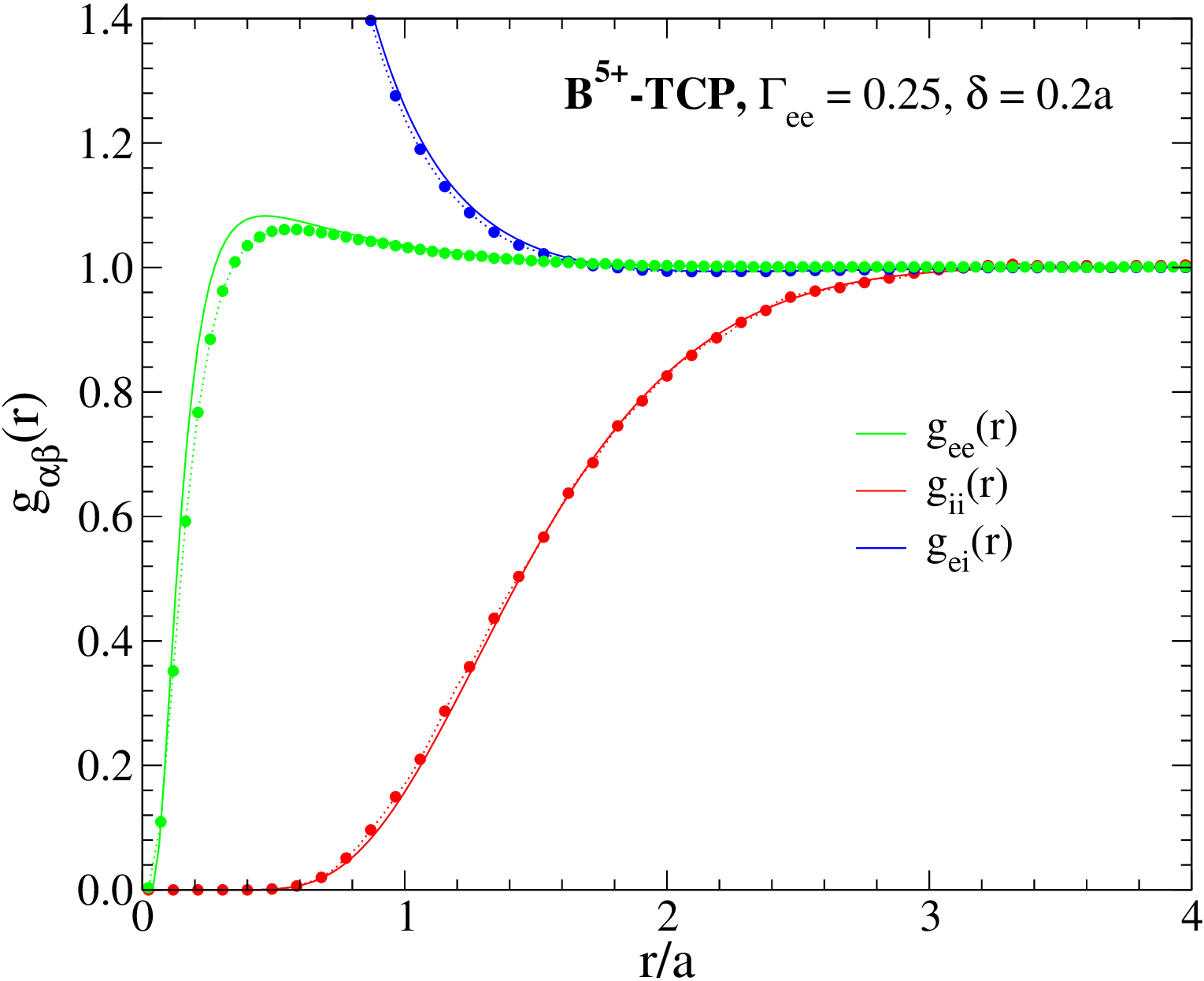}
\includegraphics[width=8cm]{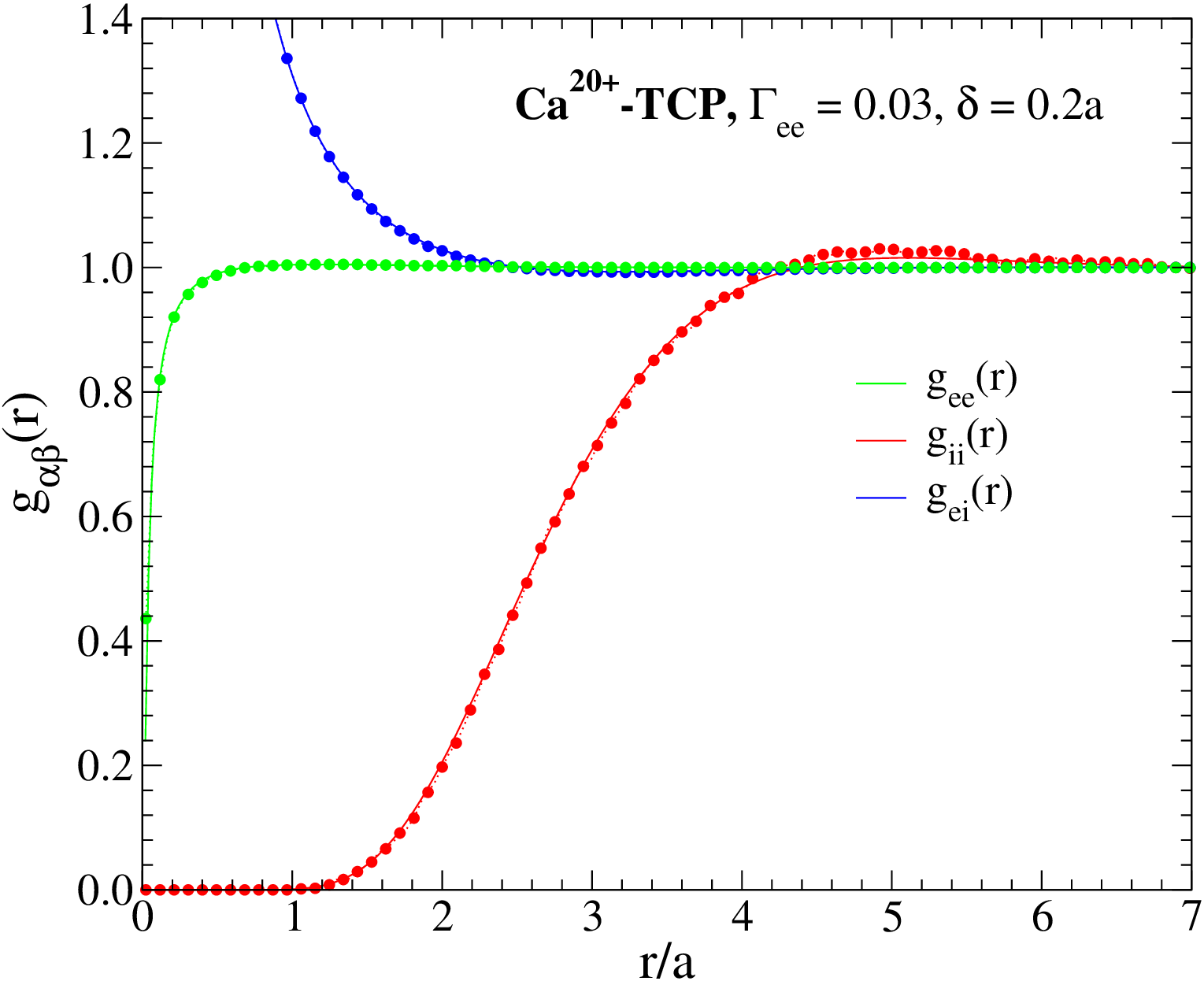}
\caption{RDFs $g_{\alpha\beta}(r)$ for B$^{5+}$ (left panel) and Ca$^{20+}$
(right panel) plasmas with $\delta =0.2a$, $\Gamma_{ee}=0.25$ and $\Gamma_{ee}=0.03$,
respectively. The lines correspond to the HNC approximation while the symbols
denote the MD simulations. The different lines represent $g_{ee}$ (green),
$g_{ii}$ (red) and $g_{ei}$ (blue).}
\label{fig:grp1}
\end{figure}

\begin{figure}[tbp]
\includegraphics[width=8cm]{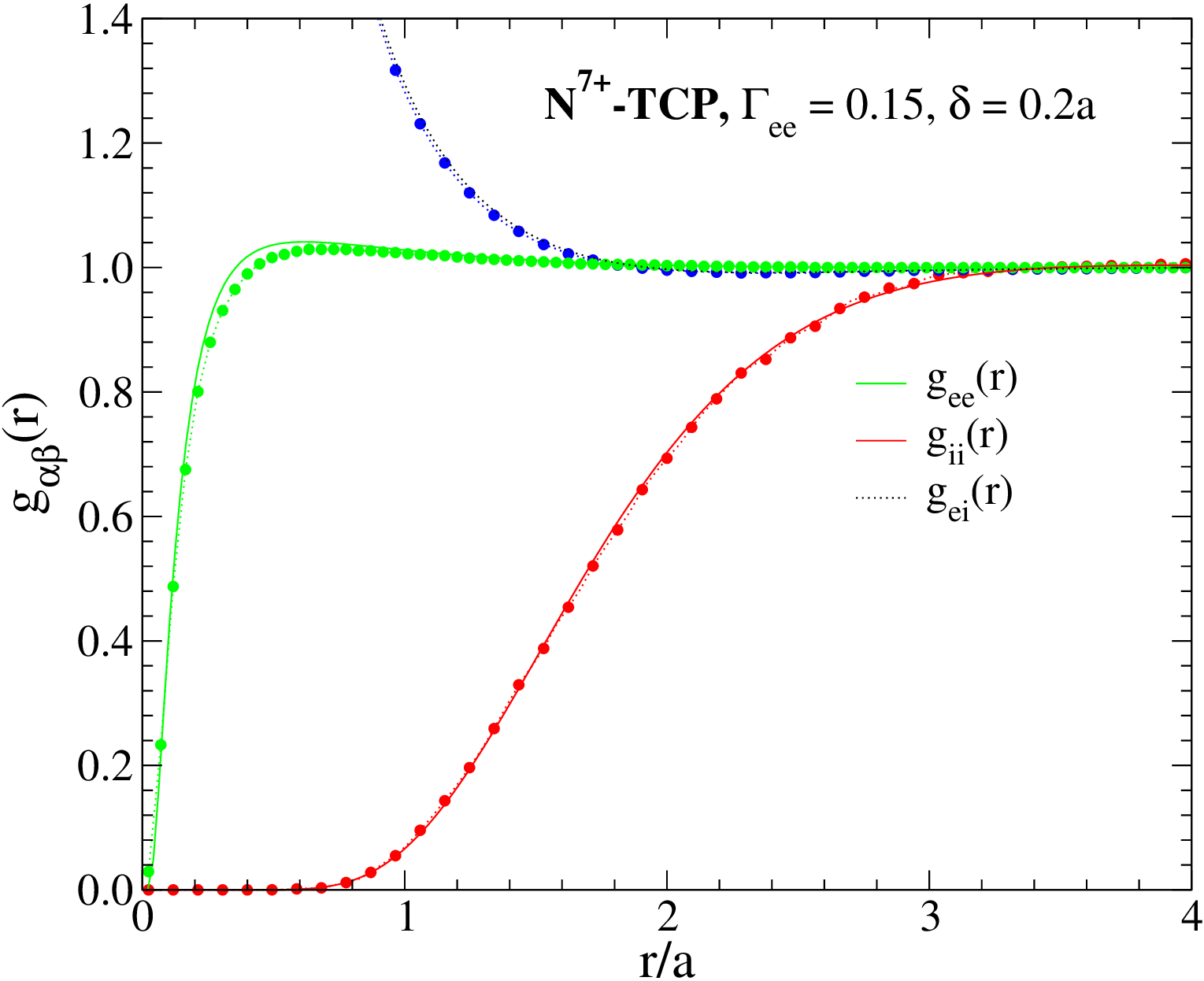}
\includegraphics[width=8cm]{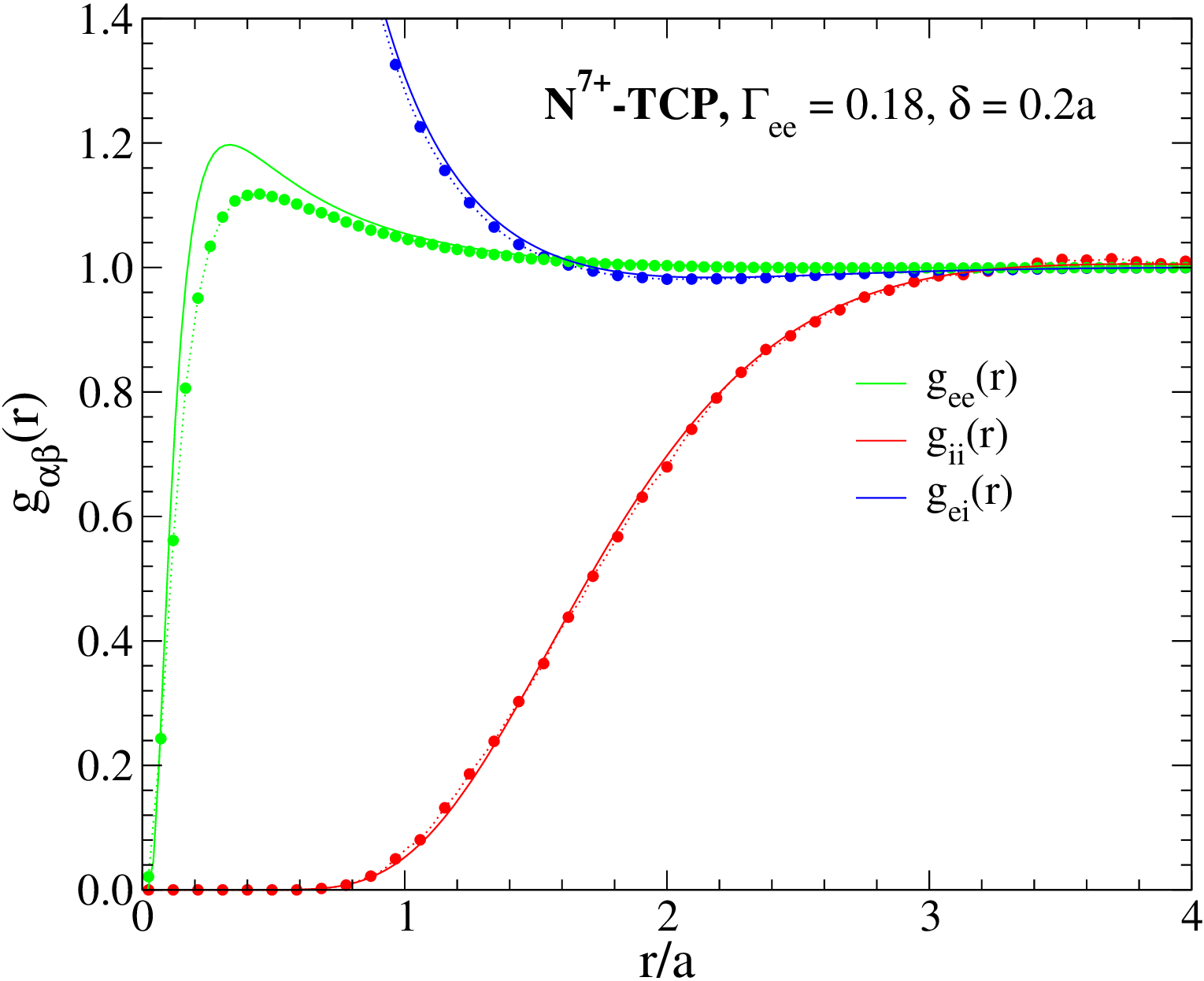}
\caption{Same as Fig.~\ref{fig:grp1} for N$^{7+}$ plasma with $\delta =0.2a$,
$\Gamma_{ee}=0.15$ (left panel) and $\Gamma_{ee}=0.18$ (right panel).}
\label{fig:grp2}
\end{figure}

\begin{figure}[tbp]
\includegraphics[width=8.5cm]{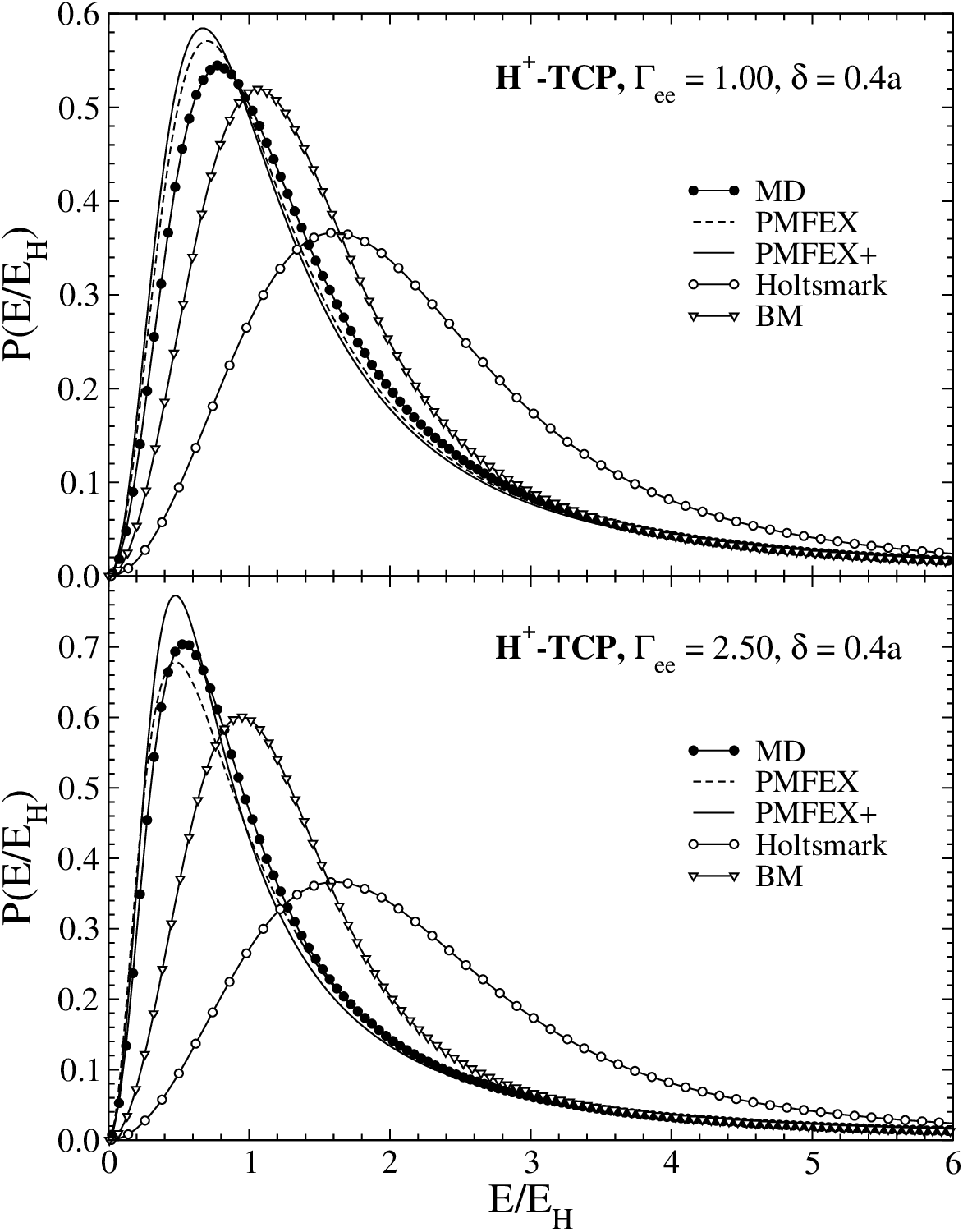}
\caption{Normalized electric microfield distributions for a hydrogen plasma
with $\delta =0.4a$, $\Gamma _{ee}=\Gamma _{ii}=1$, $\sigma_{ei} =3.15$ (top)
and $\Gamma _{ee}=\Gamma _{ii}=2.5$, $\sigma_{ei} =7.87$ (bottom)
as a function of the electric field in units of $E_{H}$. The
dashed and the solid curves are the results of the PMFEX and PMFEX+,
respectively. The filled circles represent the MFD from the MD simulations.
The Holtsmark and standard Baranger-Mozer distributions are also shown as
open circles and triangles, respectively.}
\label{fig:mfd1}
\end{figure}

\begin{figure}[tbp]
\includegraphics[width=8.5cm]{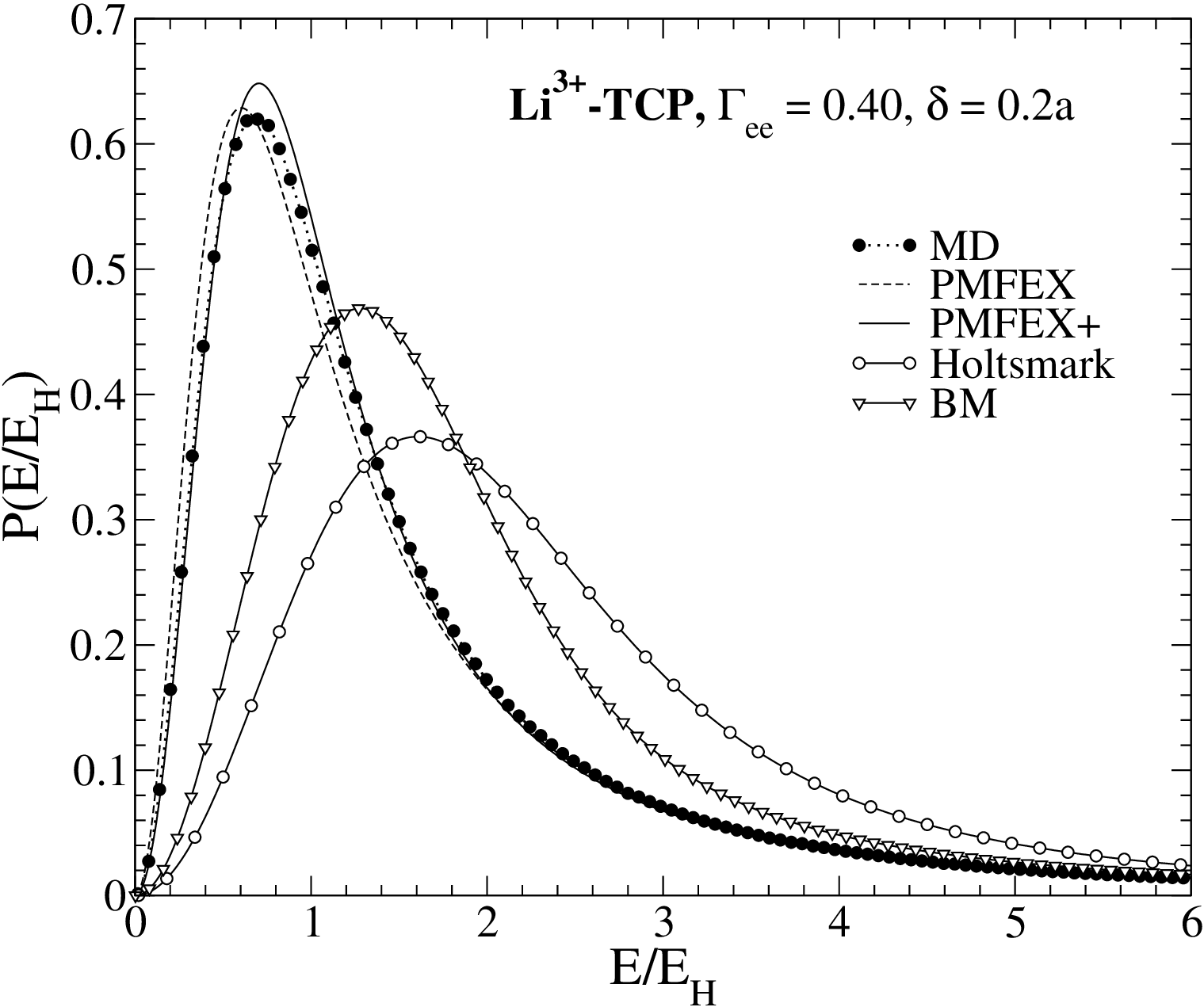}
\caption{Same as Fig.~\ref{fig:mfd1} for Li$^{3+}$ plasma with $\Gamma
_{ee}=0.4$, $\Gamma _{ii}=2.5$, $\sigma_{ei} =6.6$ and $%
\delta =0.2a$.}
\label{fig:mfd2}
\end{figure}

Before discussing the MFD itself, we first consider briefly the RDFs and 
the validity of the HNC approximation for the parameters at hand.
For small and moderate coupling where the parameter $\sigma _{ei}$ is 
well below the critical value $\sigma_{c}$, the RDFs as either calculated by 
the HNC scheme or extracted from the MD simulations are always in perfect
agreement (see e.g.~the corresponding examples given in \cite{ner05}).
Some typical examples for the RFDs at rather large coupling, where some 
deviations between HNC and MD may show up, 
are given in Figs.~\ref{fig:grp1} and \ref{fig:grp2}. 
Due to the regularization of the ion-electron interaction the
RDF $g_{ei}(r)$ is finite in the limit $r\rightarrow 0$ (not visible in
Figs.~\ref{fig:grp1} and \ref{fig:grp2}) and can be approximated as $g_{ei}(0)%
\simeq \exp(\sigma_{ei}/\mathcal{R})$ with $\mathcal{R} =1+(\delta/a)(3%
\Gamma_{ei})^{1/2}$, see Refs.~\cite{tal02,ner05}. Thus the RDF $g_{ei}(r)$
shows the expected growth of correlations with increased coupling and
decreased regularization parameter, where the HNC scheme tends to 
overestimate the value of $g_{ei}(0)$ (not visible in the figures).
For example, for N$^{7+}$ ions
$g^{\mathrm{HNC}}_{ei}(0)\simeq 57.4$, $g^{\mathrm{MD}}_{ei}(0)\simeq 48.8$
and $g^{\mathrm{HNC}}_{ei}(0)\simeq 107.3$, $g^{\mathrm{MD}}_{ei}(0)\simeq 96.4$
with $\delta =0.2a$, $\Gamma_{ee}=0.15$ and $\Gamma_{ee}=0.18$, respectively.
At strong coupling deviations also occur in the electron-electron
RDF $g_{ee}(r)$ at small $r$ (as in the case of B$^{5+}$ and N$^{7+}$). 
The strong electron-ion interaction increases
the electron density around the highly charged ion which introduces additional
correlations between the electrons. This increases the probability of close
electronic distances and results in the maxima in $g_{ee}(r)$ at distances
$r\lesssim a$. This effect is obviously again overestimated in the HNC approach.

\begin{figure}[tbp]
\includegraphics[width=8.5cm]{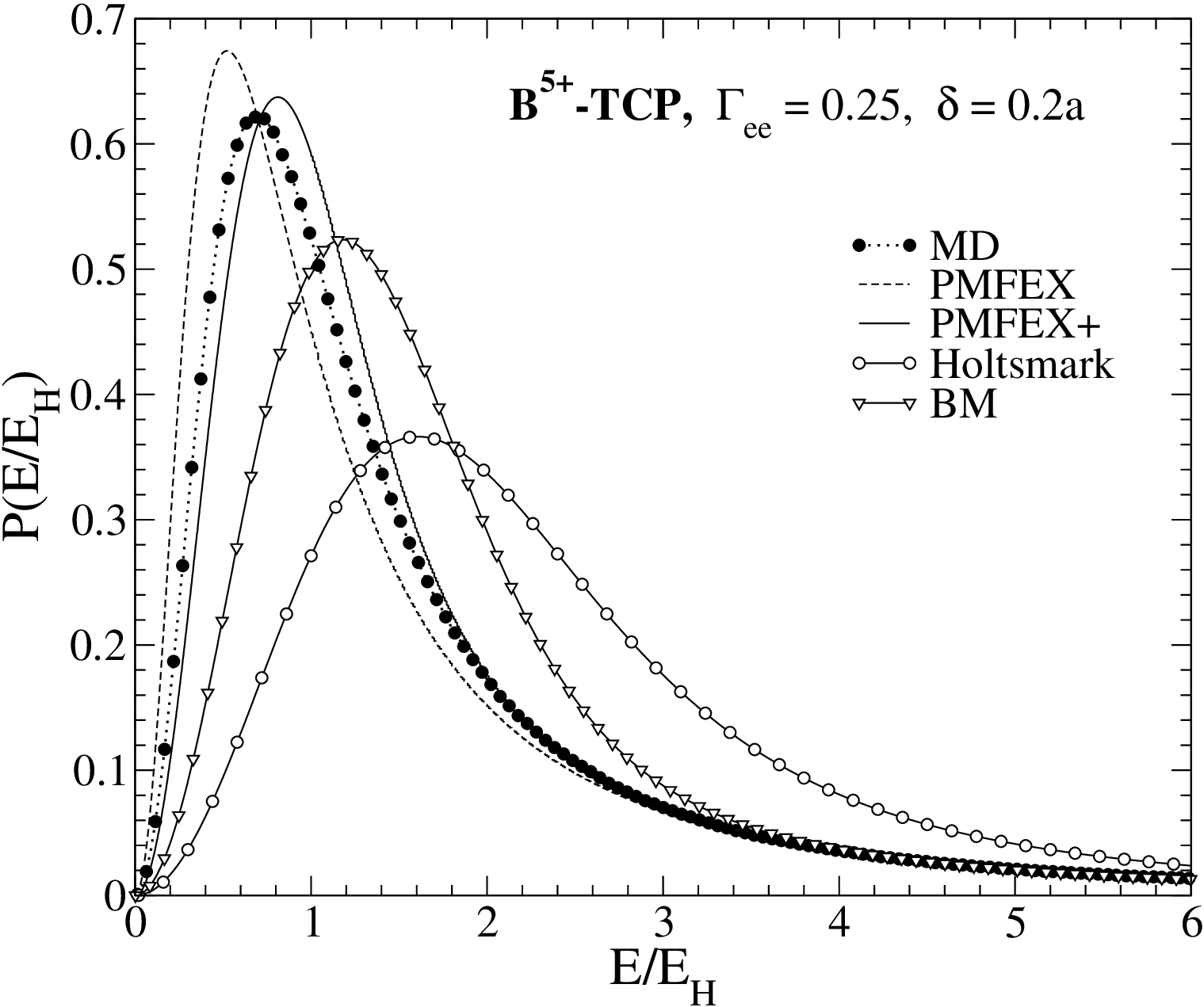}
\caption{Same as Fig.~\ref{fig:mfd1} for B$^{5+}$ plasma with $\Gamma
_{ee}=0.25$, $\Gamma _{ii}=3.66$, $\sigma_{ei} =6.64$ and $%
\delta =0.2a$.}
\label{fig:mfd3}
\end{figure}

\begin{figure}[tbp]
\includegraphics[width=8.5cm]{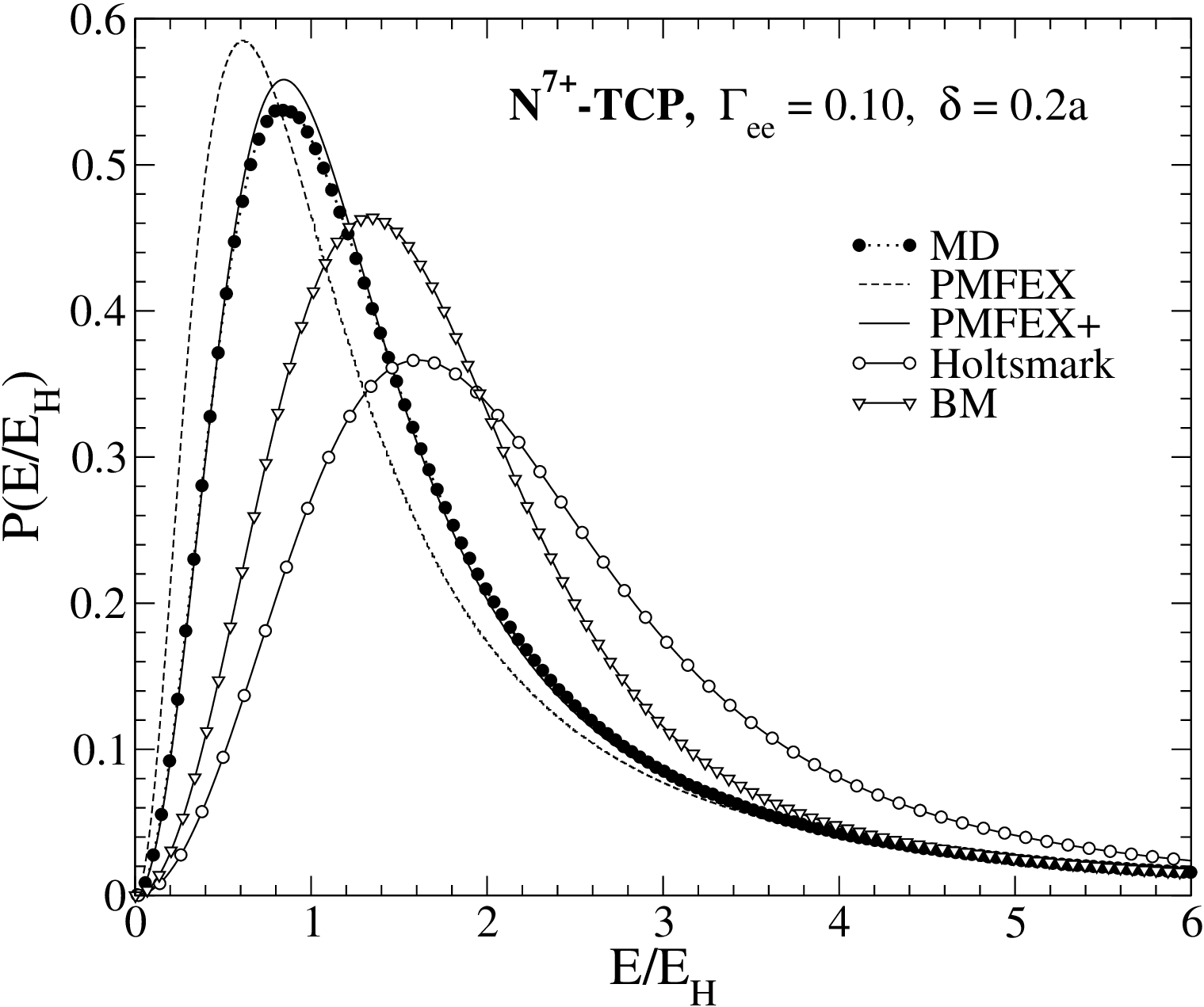}
\caption{Same as Fig.~\ref{fig:mfd1} for N$^{7+}$ plasma with $\Gamma
_{ee}=0.1$, $\Gamma _{ii}=2.56$, $\sigma_{ei} =3.66$ and $%
\delta =0.2a$.}
\label{fig:mfd4}
\end{figure}

\begin{figure}[tbp]
\includegraphics[width=8.5cm]{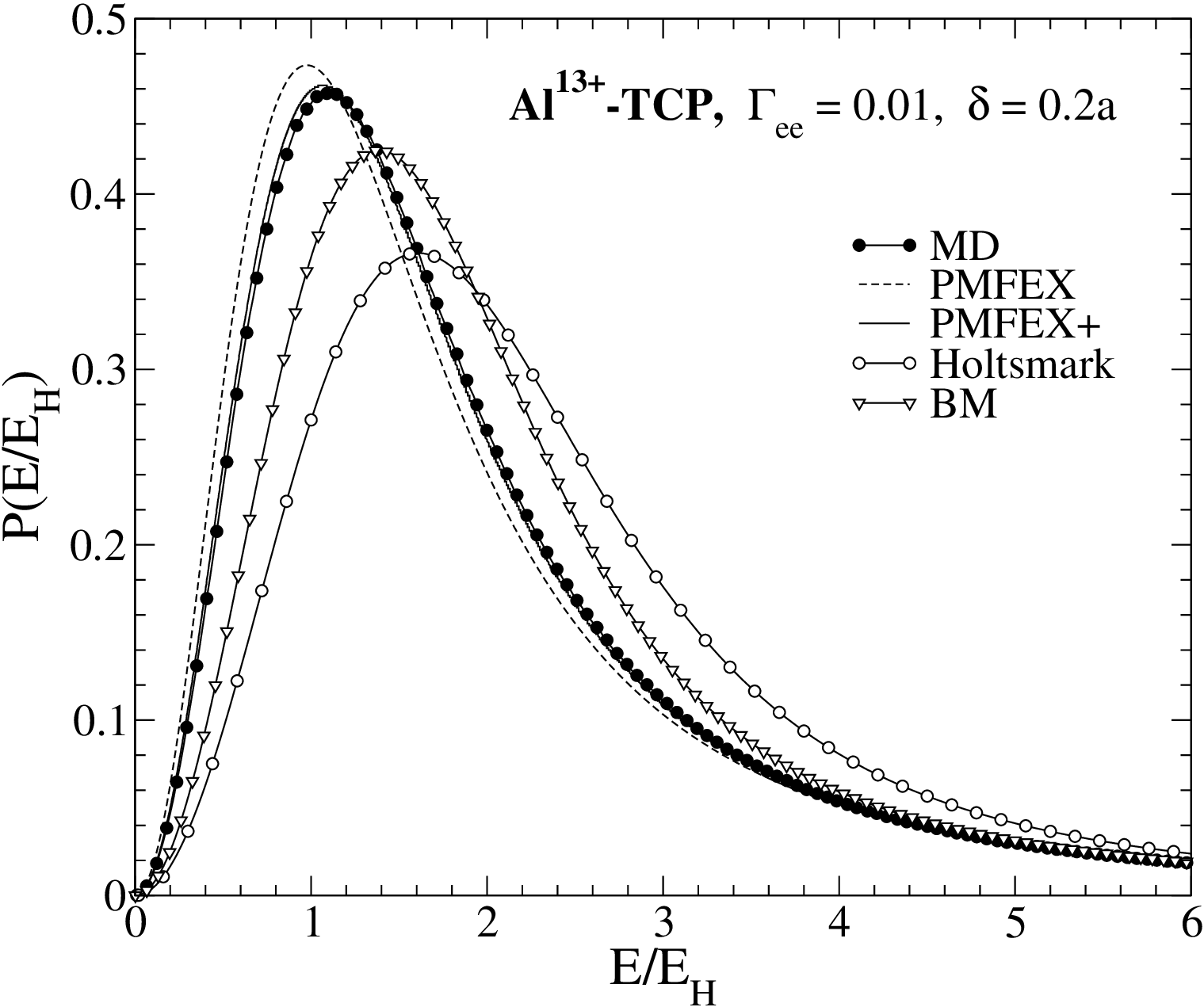}
\caption{Same as Fig.~\ref{fig:mfd1} for Al$^{13+}$ plasma with $\Gamma
_{ee}=0.01$, $\Gamma _{ii}=0.72$, $\sigma_{ei} =0.67$ and $%
\delta =0.2a$.}
\label{fig:mfd5}
\end{figure}

\begin{figure}[tbp]
\includegraphics[width=8.5cm]{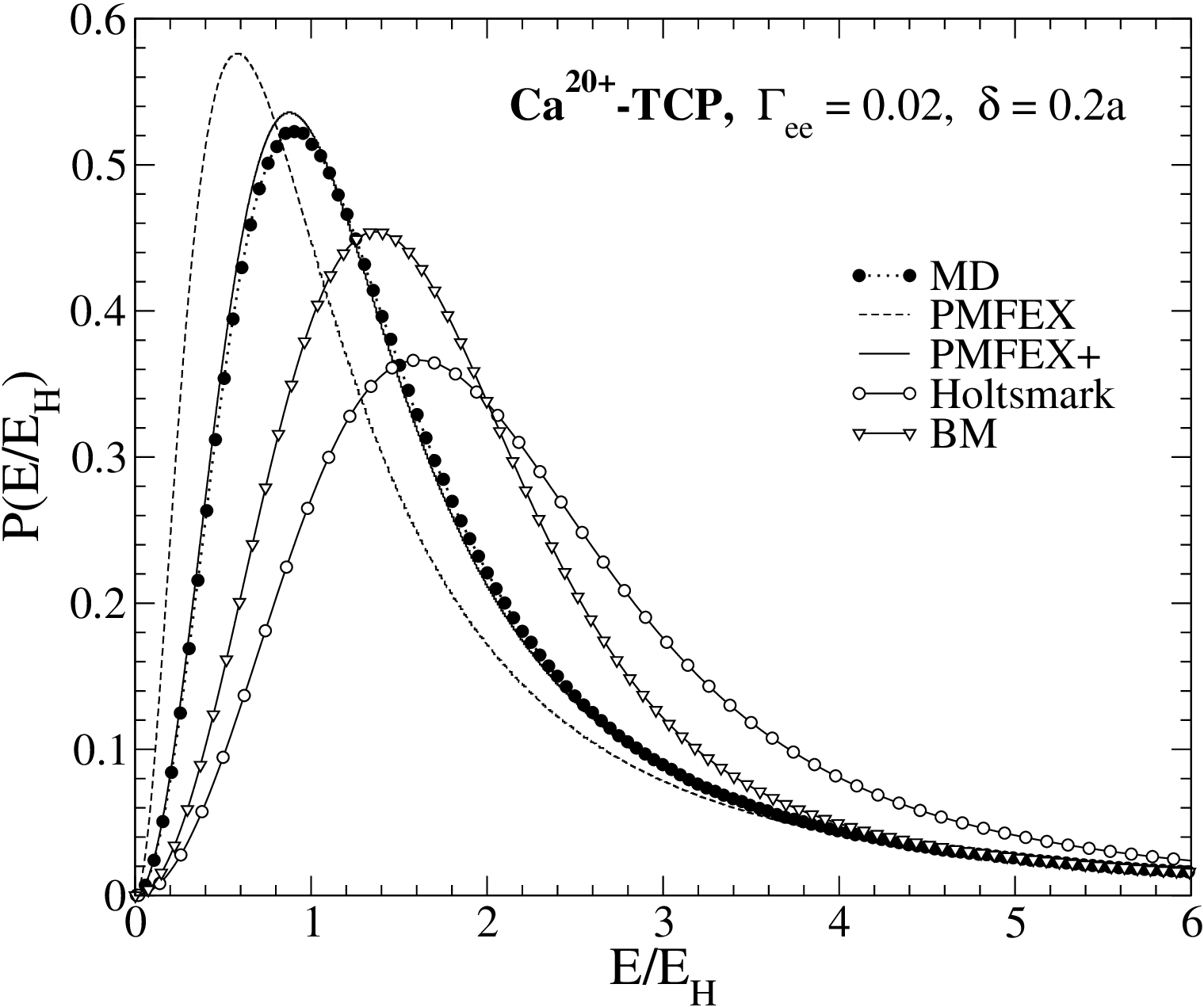}
\caption{Same as Fig.~\ref{fig:mfd1} for Ca$^{20+}$ plasma with $\Gamma
_{ee}=0.02$, $\Gamma _{ii}=2.95$, $\sigma_{ei} =2.03$ and $%
\delta =0.2a$.}
\label{fig:mfd6}
\end{figure}

In Figs.~\ref{fig:mfd1}-\ref{fig:mfd6} we next compare the MFDs calculated from the
PMFEX (dashed lines) and PMFEX+ (solid lines) schemes as well as from MD
simulations (filled circles) where the electric microfields are scaled in
units of the Holtsmark field $E_{H}$ [see Eq.~(\ref{eq:H})]. The open
circles are the Holtsmark MFDs for a TCP with Coulomb potential and 
the MFDs predicted by the first two terms of the standard BM series
Eqs.~(\ref{eq:17})-(\ref{eq:23}) are also shown as triangles. 
We here focus on cases of strong coupling where the PMFEX and PMFEX+ results 
significantly differ, i.e. where the higher order corrections
$\Delta \mathcal{L}\left( \kappa \right)$ (\ref{eq:44}) substantially 
contribute to the generating function 
$\mathcal{L}\left( \kappa \right) =\mathcal{L}_{0}\left( \kappa \right) +%
\Delta \mathcal{L}\left( \kappa \right)$.
As can be seen in all the presented cases the MFDs obtained both from the 
PMFEX and the PMFEX+ treatment strongly differ from the standard 
BM electric field distributions.

\begin{figure}[tbp]
\includegraphics[width=8.5cm]{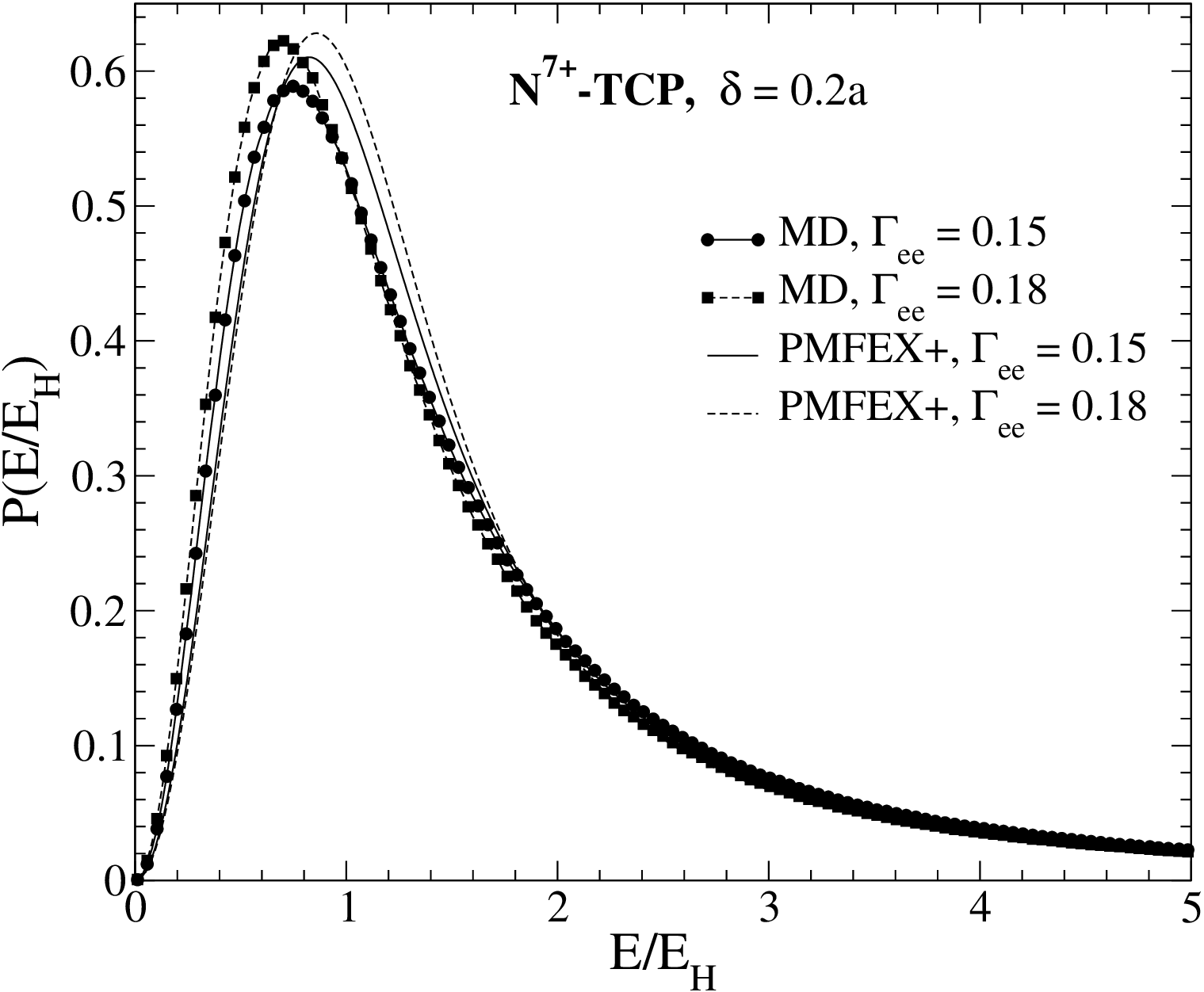}
\caption{Normalized electric microfield distributions for N$^{7+}$ plasma
with $\Gamma_{ee}=0.15$ ($\Gamma _{ii}=3.85$, $\sigma_{ei} =5.49$),
$\Gamma_{ee}=0.18$ ($\Gamma_{ii}=4.61$, $\sigma_{ei} =6.59$), and $\delta =0.2a$
as a function of the electric field in units of $E_{H}$. The solid and
dashed curves are the results of the PMFEX+ model. The filled circles and
squares represent the MFDs from the MD simulations.}
\label{fig:mfd7}
\end{figure}

\begin{figure}[tbp]
\includegraphics[width=8cm]{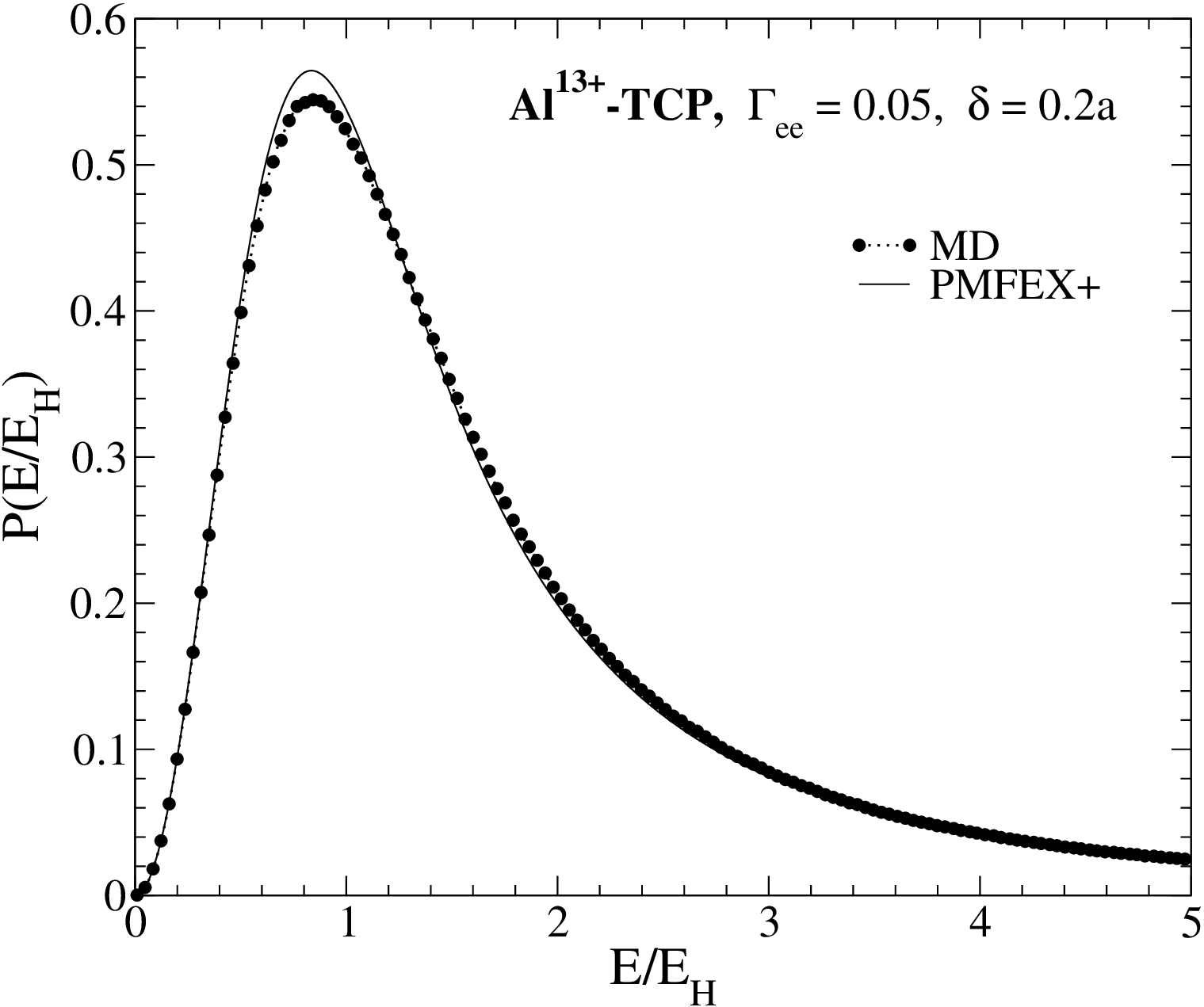}
\includegraphics[width=8cm]{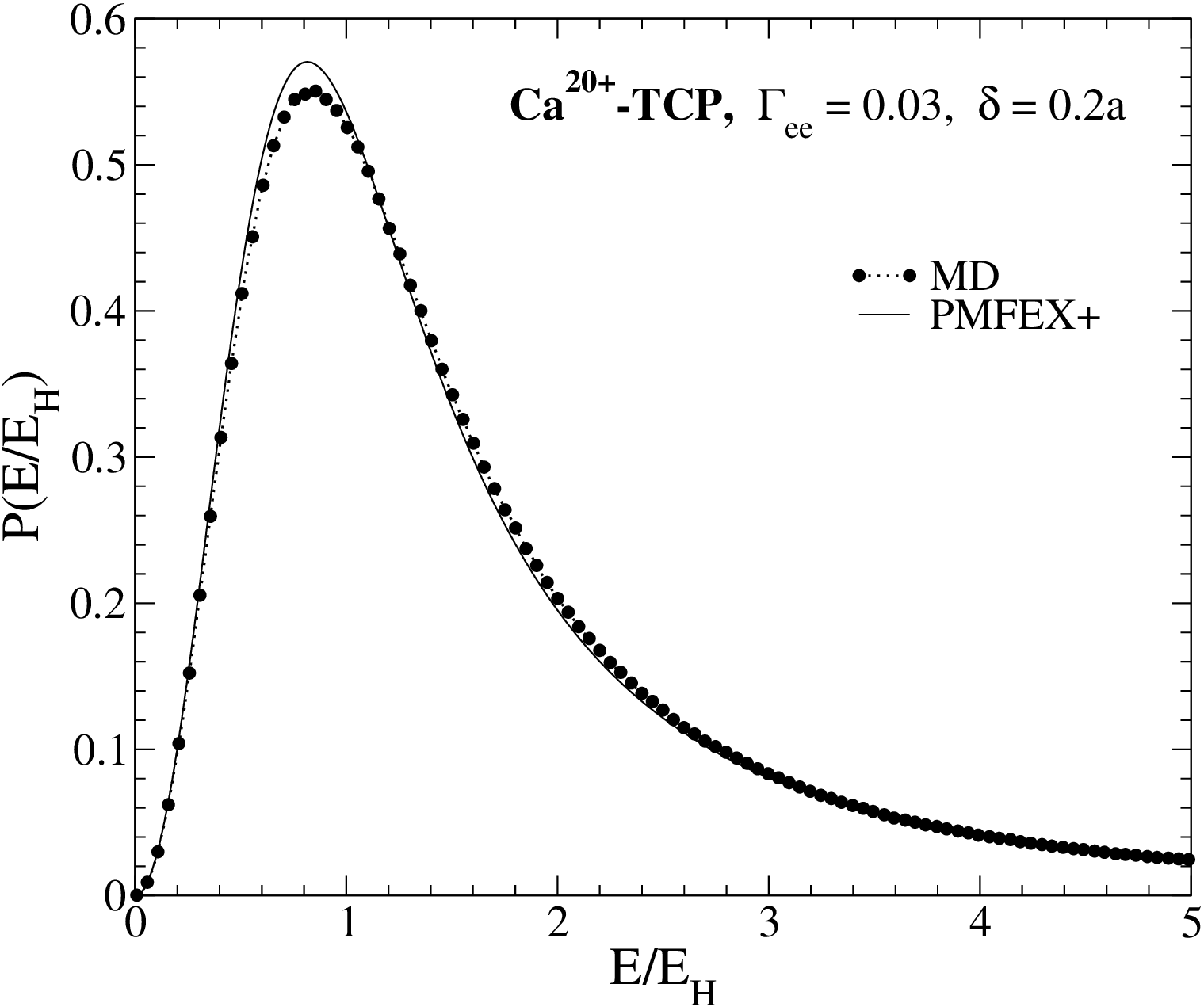}
\caption{Same as Fig.~\ref{fig:mfd7} for Al$^{13+}$ (left panel) and Ca$^{20+}$
(right panel) plasmas with $\delta =0.2a$, $\Gamma_{ee}=0.05$, $\Gamma _{ii}=3.59$,
$\sigma_{ei} =3.33$ and $\Gamma_{ee}=0.03$, $\Gamma _{ii}=4.42$, $\sigma_{ei} =3.05$,
respectively.}
\label{fig:mfd8}
\end{figure}

Concerning the agreement with the MD simulations the PMFEX+ scheme turns out 
to improve the PMFEX model substantially for large charge states of the plasma 
ions. For H$^{+}$, Fig.~\ref{fig:mfd1}, no improvement can be found, and the 
agreement with the 
MD simulations remains rather unsatisfactory, although the PMFEX(+) methods 
yield a much better description of the MFD than the standard BM approach.
For Li$^{3+}$ and B$^{5+}$, Figs.~\ref{fig:mfd2} and \ref{fig:mfd3}, the 
PMFEX+ scheme much better agrees with the MD than the PMFEX, but
obviously still deserves further improvement. 
However, for ion charges $Z>5$, Figs.~\ref{fig:mfd4}-\ref{fig:mfd6}, the MFDs 
obtained from MD simulations are in excellent agreement with those predicted 
by PMFEX+ (except for some smaller values near the maxima). The rather 
large deviation of PMFEX from PMFEX+ towards a higher probability of low 
fiels (in particular in Figs.~\ref{fig:mfd4} and \ref{fig:mfd6}) 
indicates that the PMFEX applied to the field at a neutral point does not 
correctly account for perturber-perturber correlations, and, therefore, underestimates
the MFDs at large fields.
As expected for $Z_{R}=0$ all distributions shown in 
Figs.~\ref{fig:mfd1}-\ref{fig:mfd6} merge at large electric fields, 
$\eta =E/E_{H}\gtrsim 6$ with the Holtsmark distribution with the asymptotic 
behavior $P_{H}\left( \eta \right) \simeq 1.496\eta ^{-2.5}$. 
This indicates that due to electron-electron and
ion-ion Coulomb interactions with $\delta _{ee}\simeq \delta _{ii}\simeq 0$
the second moments of the MFDs shown in Figs.~\ref{fig:mfd1}-\ref{fig:mfd6} do
not exist, see e.g. Eq.~(\ref{eq:app7}) derived in Appendix \ref{sec:app1}.

To gain some more insight of the features and the range of validity of the PMFEX and
PMFEX+ models we consider the extreme regimes of Figs.~\ref{fig:mfd7} and \ref{fig:mfd8} 
with further increased coupling parameters. In the PMFEX scheme and for 
large values of $\Gamma_{ee}$ the effective fields at large distances
behave as $\mathcal{E}_{\alpha}(r)\sim \frac{1}{r}e^{-ar} \cos(br)$. Here
$a$ and $b$ are some parameters increasing with $\Gamma_{ee}$, whereby the 
involved screening length $a^{-1}$ does not necessarily coincide with the Debye length
$\lambda$. The oscillatory nature of the effective fields drastically changes
the properties of the generating function $\mathcal{L}_{0} (\kappa)$
in the PMFEX approximation (see Eq.~(\ref{eq:44})), which becomes negative at
$\kappa E_{H}\gtrsim 1$. This is not the case in APEX (as well as in PMFEX for
charged impurity ion, i.e.~for the MFD at charged point) where the effective 
field is treated in the
Debye-H\"{u}ckel form and thus is positive and decreases monotonically with $r$.
Therefore the $\kappa$-integration in Eq.~(\ref{eq:12}) diverges and hence the
PMFEX model becomes invalid for neutral radiators ($Z_{R}=0$) (and no curves 
for the PMFEX can be given in Figs.~\ref{fig:mfd7} and \ref{fig:mfd8}).
But even in this extreme regime the PMFEX+ model remains valid and 
the agreement with MD simulations is still quite good for the lower charge 
state $Z=7$ shown in Fig.~\ref{fig:mfd7} (although not as good as 
in Fig.~\ref{fig:mfd4}), and almost perfect for the highly charged ions 
as shown in Fig.~\ref{fig:mfd8}.

As discussed above some deviations
between HNC scheme and MD simulations occur in the electron-electron RDF
$g_{ee}(r)$ at small distances $r\lesssim a$ for B$^{5+}$ and N$^{7+}$ TCPs,
see the left panel of Fig.~\ref{fig:grp1} and Fig.~\ref{fig:grp2}. Similarly
the static structure factor $S_{ee}(k)$ obtained from HNC deviates from MD
data at $ka\gtrsim 1$. This may be critical for the accuracy of the electronic
effective field, Eq.~(\ref{eq:48}) and hence for the generating function
$\mathcal{L} (\kappa)$ Eqs.~(\ref{eq:42})-(\ref{eq:44}) calculated from the
HNC approximation. Since the generating function is involved in the exponential
factor $e^{-\mathcal{L} (\kappa)}$ even small deviations in $\mathcal{L} (\kappa)$
results some appreciable deviations in the MFD. Apparently the PMFEX+ model
could be improved in the case of B$^{5+}$ and N$^{7+}$ TCPs in Figs.~\ref{fig:mfd3}
and \ref{fig:mfd7}, respectively, using the MD data for the RDF $g_{ee}(r)$.

\section{Discussion and Conclusion}
\label{sec:sum}

In this paper we investigate the microfield distributions in a two-component
plasmas with attractive electron-ion interactions. Based on the
renormalization of the standard Baranger-Mozer cluster expansion technique
our objective here was to derive the corrections to the PMFEX model proposed
recently in Ref.~\cite{ner05}. The outlined theoretical method is
abbreviated as PMFEX+. In general the MFDs predicted by PMFEX for the case
of charged impurity ion are very accurate and its corrections may be quite
small, except the case of strongly coupled hydrogen at small
regularization parameter $\delta $ \cite{ner05}. As discussed in Ref.~\cite%
{ner06} the PMFEX is less accurate in the case of the neutral point and 
an improvement of the model is required. We thus focused on testing the
predictions of the PMFEX+ approximation for the neutral point based on the
HNC treatment of static correlations by confronting it with the MFDs
obtained from MD simulations. 
One of the basic assumptions made for all these models is the regularization 
of the attractive electron-ion interaction
at short distances to introduce quantum diffraction effects in the employed
classical approach. For the repulsive electron-electron and ion-ion 
interactions we assumed a bare Coulomb potential, for simplicity.

As examples the  cases of H$^{+}$, Li$^{3+}$, B$^{5+}$, N$^{7+}$,
Al$^{13+}$ and Ca$^{20+}$ two-component plasmas (TCPs) with symmetric and
largely asymmetric densities of the plasma species were considered. Our
treatment is limited to a parameter regime with $\sigma <\sigma _{c}$, where
the critical values $\sigma _{c}$ for $\delta =0.1a$, $0.2a$ and $0.4a$ are
shown in Table~\ref{tab:1}. Within this parameter regime the $g_{\alpha
\beta }(r)$ from the HNC equations agree well with the MD simulations
(except the correlation function $g_{ee}(r)$ for the cases given in
Figs.~\ref{fig:grp1} and \ref{fig:grp2}). Beyond these critical $\sigma $
the HNC equations do either not converge or end up in unphysical solutions.
A further increase of the coupling parameters also leads to the formation of
classical bound electronic states in the MD simulations with no corresponding
quantum counterpart.

For high ion charges $Z>5$, like N$^{7+}$, Al$^{13+}$ and Ca$^{20+}$ the
PMFEX model is substantially improved and the agreement of the PMFEX+ model
with the MD simulations is excellent, as shown in Figs.~\ref{fig:mfd4}-\ref{fig:mfd6}
and \ref{fig:mfd8}. For a TCP with light ions, e.g. the case of Li$^{3+}$
ions in Fig.~\ref{fig:mfd2}, the PMFEX and PMFEX+ differ, but both somewhat
deviate from the MD simulations as in the case of H$^{+}$ TCP. In this regime
with moderate ionic charge state $Z$ and for large coupling the renormalized
Baranger-Mozer series should probably be truncated at higher orders. We
have also shown that with increasing coupling the PMFEX scheme becomes invalid
while the PMFEX+ approach still works and yields good or even almost perfect 
agreement with the MD, see Figs.~\ref{fig:mfd7} and \ref{fig:mfd8}.

In summary we note that the PMFEX approximation incorporates some key
features of the MFD, in particular for large microfields $E$ (small $\kappa $
values). The large fields are predominantly due to configurations with a
single-plasma particle near the impurity ion. In this case the
perturber-perturber correlations are less important and the PMFEX as an
independent-particle model is accurate for both neutral and charged
radiators. The behavior of the MFD at large $E$ is closely related to the
behavior of the generating function $\mathcal{L}\left( \kappa \right) $ at
small $\kappa $ and how much accurate the exact second moment is involved in
the model. At small $\kappa $ the generating function behaves as $\mathcal{L}%
\left( \kappa \right) \simeq (\kappa ^{2}/6)\langle E^{2}\rangle $\ (see
Eq.~(\ref{eq:14})), where the second moment $\langle E^{2}\rangle $ is
exactly involved in the PMFEX model. Furthermore the microfields in the
large field regime are asymptotically Gaussian distributed and characterized
by the second moment $\langle E^{2}\rangle $. As discussed above this is,
however, not so for a neutral radiator and assuming bare Coulomb interactions 
between plasma particles where $\mathcal{L}\left( \kappa \right) $ behaves as $\sim
\kappa ^{3/2}$ at small $\kappa $ and the MFDs is similar to the Holtsmark
distribution at large $E$. Conversely, the small microfields (large $\kappa 
$\ values) are due to the additive effects of many particles at large
distances and the PMFEX approximation is again accurate \cite{ner05}
except in the case of the neutral point where some deviations from MD
simulations may occur, see Figs.~\ref{fig:mfd1}-\ref{fig:mfd6} and 
Ref.~\cite{ner06} for other examples. The main uncertainty in PMFEX, therefore,
remains the domain of the intermediate configurations with $E\sim 1$ and $%
\kappa \sim 1$ (in units of $E_{H}$ and $1/E_{H}$, respectively).
Apparently, this domain is reduced by means of the renormalized (screened)
fields $\mathcal{E}_{\alpha }\left( r\right) $ and the PMFEX is much more
adequate compared to the standard Baranger-Mozer (second order) treatment.
In PMFEX+ model, on the other hand, imposing the renormalization procedure
to the standard Baranger-Mozer series substantially improves their
convergency, since $\psi _{a}^{(\alpha )}(\boldsymbol{\kappa })$ tends to
zero for the relevant configurations more rapidly than $\chi _{a}^{(\alpha
)}(\boldsymbol{\kappa })$. Finally, the PMFEX+ model may also be useful for
a charged impurity ion in the cases where the PMFEX approximation deviates from
MD. For the TCP, an example is the case of strongly coupled hydrogen with
small regularization parameter $\delta $ considered in Ref.~\cite{ner05}.

\begin{acknowledgments}
This work was supported by the Bundesministerium f\"{u}r Bildung und
Forschung (BMBF, 06ER145) and by the Gesellschaft f\"{u}r
Schwerionenforschung (GSI, ER/TOE). The work of H.B.N. and D.A.O. has been
partially supported by the Armenian Ministry of Higher Education and Science
Grant No.~0247. We gratefully acknowledge many stimulating discussions with
C. Toepffer.
\end{acknowledgments}

\appendix

\section{Exact second moment for neutral radiator}
\label{sec:app1}

In this Appendix we derive the exact second moment for the MFD in the TCP at
the neutral point. We use the exact relation for the second moment derived
in Ref.~\cite{ner05} 
\begin{equation}
\left\langle E^{2}\right\rangle =\frac{k_{B}Tn_{e}}{Z_{R}\varepsilon _{0}}%
\left( \int_{0}^{\infty }\widetilde{u}_{e}\left( r\right) g_{eR}\left(
r\right) dr-\int_{0}^{\infty }\widetilde{u}_{i}\left( r\right) g_{iR}\left(
r\right) dr\right) ,  \label{eq:app1}
\end{equation}%
where $\widetilde{u}_{\alpha }\left( r\right) =-\left[ r^{2}u_{\alpha
R}^{\prime }\left( r\right) \right] ^{\prime }$ with $\alpha =e,i$\ and the
pair interaction potentials $u_{\alpha R}\left( r\right) $ are given by Eq.~(%
\ref{eq:3}) with the regularization parameters $\delta _{\alpha R}$.

The second moment $\langle E^{2}\rangle $ is ill-defined for the
neutral-point distribution since $Z_{R}\rightarrow 0$ and $g_{\alpha
R}\left( r\right) \rightarrow 1$ and the expression in the brackets in Eq.~(%
\ref{eq:app1}) vanishes in this case (this is true for the Coulomb potential
as well as for any potential regularized at the origin). To obtain the
correct limit of Eq.~(\ref{eq:app1}) at $Z_{R}\rightarrow 0$ we recall the
definition of the correlation functions $g_{\alpha R}\left( r\right) $ \cite%
{ner05}. At vanishing $Z_{R}$\ this functions read 
\begin{equation}
g_{\alpha R}(r)-1=-\frac{Z_{R}e_{S}^{2}}{k_{B}T}\left\{ q_{\alpha }u_{\alpha
R}\left( r\right) +\sum_{\beta }q_{\beta }n_{\beta }\int d\mathbf{r}%
_{1}u_{\beta R}\left( r_{1}\right) \left[ g_{\alpha \beta }\left( \left\vert 
\mathbf{r}-\mathbf{r}_{1}\right\vert \right) -1\right] \right\} +\mathrm{O}%
\left( Z_{R}^{2}\right) .  \label{eq:app2}
\end{equation}%
Here $g_{\alpha \beta }\left( r\right) $ are the equilibrium correlation
functions. Note that the second term in the right-hand side of Eq.~(\ref%
{eq:app2}) is the excess potential energy of the TCP. In particular, using
Eq.~(\ref{eq:app2}) it is straightforward to calculate the effective field $%
\mathcal{E}_{\alpha }\left( r\right) $ for the neutral point. Insertion of
this relation into Eq.~(\ref{eq:30}) in the limit $Z_{R}\rightarrow 0$\ and
for Coulomb electron-electron, ion-ion interactions yield Eq.~(\ref{eq:29})
with $g_{\alpha R}\left( r\right) =1$ (see also Eq.~(\ref{eq:48})).

Now we substitute Eq.~(\ref{eq:app2}) into Eq.~(\ref{eq:app1}). In the limit
of vanishing charge $Z_{R}\rightarrow 0$ this yields 
\begin{eqnarray}
\left\langle E^{2}\right\rangle &=&4\pi \sum_{\alpha }n_{\alpha
}\int_{0}^{\infty }E_{\alpha }^{2}\left( r\right) r^{2}dr  \label{eq:app3} \\
&&+16\pi ^{2}n_{e}^{2}e_{F}^{2}\left\{ \sum_{\alpha }\int_{0}^{\infty }\left[
g_{\alpha \alpha }\left( r\right) -1\right] \Psi _{\alpha \alpha }\left(
r\right) rdr-2\int_{0}^{\infty }\left[ g_{ei}\left( r\right) -1\right] \Psi
_{ei}\left( r\right) rdr\right\} ,  \notag
\end{eqnarray}%
where $E_{\alpha }\left( r\right) =-q_{\alpha }e_{F}u_{\alpha R}^{\prime
}\left( r\right) $ is the single-particle electric field with $q_{e}=-1$ and 
$q_{i}=Z$, $U_{\beta }\left( r\right) =\left[ ru_{\beta R}\left( r\right) %
\right] ^{\prime }$, and 
\begin{equation}
\Psi _{\alpha \beta }\left( r\right) =\frac{1}{2}\int_{0}^{\infty }\left[
U_{\beta }\left( \left\vert \rho -r\right\vert \right) -U_{\beta }\left(
\rho +r\right) \right] u_{\alpha R}\left( \rho \right) \rho d\rho .
\label{eq:app4}
\end{equation}%
The first term in Eq.~(\ref{eq:app3}) is the averaged density of the
electric microfield energy (self-energy), the second one is the density of
the excess electric energy which appears due to the correlations between
plasma particles. Assuming regularized interaction potential Eq.~(\ref{eq:3}%
) we obtain $U_{\beta }\left( r\right) =e^{-r/\delta _{\beta R}}/\delta
_{\beta R}$, and 
\begin{eqnarray}
\Psi _{ei}\left( r\right) &=&1-\frac{1}{2}e^{-r/\delta _{iR}}-\frac{%
e^{-r/\delta _{eR}}-e^{-r/\delta _{iR}}}{2\left( 1-\Delta \right) }-\frac{%
e^{-r/\delta _{eR}}+\Delta e^{-r/\delta _{iR}}}{2\left( 1+\Delta \right) },
\label{eq:app5} \\
\Psi _{\alpha \alpha }\left( r\right) &=&1-\left( 1+\frac{r}{2\delta
_{\alpha R}}\right) e^{-r/\delta _{\alpha R}}.  \label{eq:app6}
\end{eqnarray}%
Here $\Delta =\delta _{iR}/\delta _{eR}$. In this case Eq.~(\ref{eq:app3})
yields 
\begin{eqnarray}
\left\langle E^{2}\right\rangle &=&2\pi n_{e}e_{F}^{2}\left( \frac{1}{\delta
_{eR}}+\frac{Z}{\delta _{iR}}\right)  \label{eq:app7} \\
&&+16\pi ^{2}n_{e}^{2}e_{F}^{2}\left\{ \sum_{\alpha }\int_{0}^{\infty }\left[
g_{\alpha \alpha }\left( r\right) -1\right] \Psi _{\alpha \alpha }\left(
r\right) rdr-2\int_{0}^{\infty }\left[ g_{ei}\left( r\right) -1\right] \Psi
_{ei}\left( r\right) rdr\right\} ,  \notag
\end{eqnarray}%
where $\Psi _{\alpha \alpha }\left( r\right) $ and $\Psi _{ei}\left(
r\right) $ are now given by Eqs.~(\ref{eq:app5}) and (\ref{eq:app6}). The
first term in Eq.~(\ref{eq:app7}) is precisely the second moment of the
Holtsmark distribution obtained in Ref.~\cite{ner05} for the regularized
interactions which is independent on $Z_{R}$. The second term arises due to
correlations between particles. For the Coulomb interaction between plasma
particles and the radiator with $\delta _{eR}\simeq 0$ or $\delta
_{iR}\simeq 0$ the first term in Eq.~(\ref{eq:app7}) diverges and the second
moment of the MFD does not exist in this case.

\end{document}